\documentclass[aip,amsmath,amssymb,jcp,preprint]{revtex4-1}

\draft 

\usepackage{siunitx}
\usepackage{physics}
\DeclareSIUnit\angstrom{\protect \text {Å}}
\usepackage{bm}
\usepackage{upgreek}
\usepackage{multirow}
\usepackage{booktabs}
\usepackage{cleveref}
\usepackage[version=4]{mhchem}
\usepackage{comment}
\usepackage[pdftex]{graphicx}

\crefname{section}{Section}{Sections}
\crefname{subsection}{Subsection}{Subsections}
\crefname{equation}{Eq.}{Eqs.}
\crefname{figure}{Fig.}{Figs.}
\crefname{table}{Table}{Tables}
\crefname{appendix}{Appendix}{Appendices}

\begin{document}


\title{Atomic momentum distributions in polyatomic molecules in rotational-vibrational eigenstates} 



\author{Sota Sakaguchi}

\author{Yasuhiro Ohshima}

\author{Masakazu Yamazaki}
\email[]{yamazaki@chem.titech.ac.jp}
\affiliation{Department of Chemistry, School of Science, Tokyo Institute of Technology, 2-12-1, Ookayama, Megro-ku, Tokyo 152-8550, Japan}


\date{\today}

\begin{abstract}
We report a quantum mechanical method for calculating the momentum distributions of constituent atoms of polyatomic molecules in rotational-vibrational eigenstates. 
Application of the present theory to triatomic molecules in the rovibrational ground state revealed that oscillatory changes appear on the proton momentum distribution in the nonlinear \ce{H2O} molecule, whilst no such modulation is present in the case of an oxygen atom in the linear \ce{CO2} molecule. 
The atomic momentum distributions were analyzed in detail by means of a rigid rotor model, and it was found that the oscillation originates from quantum-mechanical delocalization of the target atom with respect to the other atoms.
\end{abstract}

\pacs{}

\maketitle 

\section{Introduction}
\label{Introduction}

Nuclear quantum effects (NQEs), such as zero-point energy and tunneling, play a crucial role in numerous phenomena of importance in materials science, chemistry, and biology, e.g., static and dynamical properties of hydrogen bonding systems, proton transfer reactions, and diffusion in bulk and confined environments.\cite{ceriotti2016nuclear,tuckerman2018preface,fang2019quantum} 
Deep Inelastic Neutron Scattering (DINS)\cite{evansDeepInelasticNeutron1993,watson1996neutron,mayers2002Measurement,andreani2005measurement,andreani2017electronvolt} has attracted recently considerable interest, because it provides information on momentum distribution of constituent atoms in matter, which is sensitive to the NQEs. 
For example, DINS experiments have revealed a striking bimodal profile of the proton momentum distribution in KDP,\cite{reiter2002direct} confined water in nanomaterials,\cite{reiter2012evidence,garbuio2007proton} and hydrated water in proteins.\cite{senesi2007proton}
The bimodal shape was regarded as a result of the proton tunneling, or coherent delocalization of the proton over different sites.\cite{reiter2004proton,garbuio2007proton} 
Furthermore, the oscillatory behavior of the atomic momentum distribution has been also reported for a tunneling phenomenon with respect to rotational motion of water in beryl.\cite{kolesnikov2016quantum}

Many of the DINS experimental results, including the bimodal distribution described above, were interpreted by means of an effective single-particle model, where atomic momentum distributions is evaluated under approximation with single-particle motion in an effective (Born-Oppenheimer) mean field potential.\cite{reiter2002direct,reiter2004proton,reiter2012evidence,senesi2007proton,garbuio2007proton,reiter1985measurement,andreani2001single,homouz2007measurement, flammini2012sphericala}
However, it remains uncertain whether or not the correlated motions in quantum many-body systems can be reduced to the single-particle motions, and the interpretation of the DINS experimental results based on the single-particle model is subject to some controversy.\cite{soper2009comment,lin2011momentum,wu2020quantum}
In order to elucidate the physics behind the atomic momentum distribution, an associated quantum calculation that treats all degrees of freedom of nuclei are of crucial importance.

Path integral molecular dynamics (PIMD) is the state-of-the-art method for calculating atomic momentum distributions in molecular systems.\cite{lin2011momentum,wu2020quantum,morrone2008nuclear,lin2010displaced,ceriotti2012efficient}
Using PIMD, ensemble averages of atomic momentum distributions can be calculated under explicit consideration of quantum behavior of nuclei (except for particle exchange) in many-body systems.
Although PIMD results can be compared directly to experimental results, it is difficult to capture individual contributions from  an each single quantum state, since they are smeared out by the ensemble average.\cite{wu2020quantum}
To investigate such NQEs at the most fundamental level, it is essential to develop a new theory suitable to a single quantum state of isolated molecules, whose wavefunctions are usually well characterized. 

For diatomic molecules in their rotational-vibrational eigenstates, such a quantum theory for the atomic momentum distribution has already been well established.\cite{colognesi1999deep,andreani1995deepa,tachibana2022direct}
Colognesi \textit{et al.} revealed that the atomic momentum distribution exhibits a clear oscillation originating from the quantum-mechanical rotation, of which wavelength is related to the internuclear distance.\cite{colognesi1999deep}  
To discuss the generality of this oscillation for three-atom or larger systems, it is highly expected to extend the quantum treatment from diatomic molecules to any kind of polyatomic molecules.

At present, however, the quantum effects on atomic momentum distributions in polyatomic molecules remains still to be investigated. 
Colognesi \textit{et al.} have proposed a quantum treatment based on the position autocorrelation function, but the target is restricted to symmetric top molecules.\cite{colognesi2001deep}
They calculated the proton momentum distribution of the ground state \ce{NH3} but did not investigate any quantum effect, by simply concluding that molecular rotations can be dealt with classically at higher temperature.
Thus, issue still remains; whether or not the quantum oscillation is also present in the atomic momentum distribution in any kind of molecules other than diatomics; if it presents, what is the origin of it? 

In this study, we report a quantum treatment for calculating the atomic momentum distribution in polyatomic molecules in rotational-vibrational eigenstates.
Different from the previous method based on the autocorrelation function,\cite{colognesi2001deep} our approach utilizes momentum-space molecular wavefunctions themselves, which is a significant extension of the quantum treatment for diatomic molecules\cite{colognesi1999deep} to polyatomic molecules having arbitrary symmetry.
A practical example of triatomic molecules (\ce{H2O} and \ce{CO2}) in its rovibrational ground state has clearly shown a distinct oscillation appeared in the proton momentum distribution in the nonlinear \ce{H2O} molecule but not in the case of an oxygen atom in the linear \ce{CO2} molecule. 
The origin of the oscillatory structure and the difference in the results between \ce{H2O} and \ce{CO2} are discussed in terms of the quantum-mechanical distribution of the target atom with respect to the other constituent atoms, which is determined by the rotational-vibrational wavefunction.

\section{Theory}
\subsection{Atomic momentum distributions obtained through momentum-space
 molecular wavefunctions}
The momentum distribution $n_s$ of the $s$th atom in an \textit{N}-atomic molecule in a single quantum state is defined by
\begin{align}
    n_s(\vb{P}_s)=
    &\int\dd\vb{P}_1\cdots\dd\vb{P}_{s-1}\dd\vb{P}_{s+1}\cdots\dd\vb{P}_N
    \notag\\
    &\times\abs{\Phi_X(\vb{P}_1,\dots,\vb{P}_N)}^2,
    \label{eq:nn=int_abspsi^2}
\end{align}
where $\vb{P}_i=(P_{iX},P_{iY},P_{iZ})$ is the momentum of the $i$th atom, and $\Phi_X$ is the momentum-space molecular (translational, rotational, and vibrational) wavefunction in the Cartesian coordinate system $(X_a)\equiv(X_1,\dots,X_{3N})=(\vb{R}_1,\dots,\vb{R}_N)$ with the position $\vb{R}_i=(R_{iX},R_{iY},R_{iZ})=(X_{3i-2},X_{3i-1},X_{3i})$ of the $i$th atom. 
Since the momentum and position operators satisfy the canonical commutation relation, $\Phi_X$ is obtained by the Fourier transform of the position-space wavefunction $\Psi_X$ in the Cartesian coordinate system $(X_a)$,
\begin{align}
    \Phi_X(\vb{P}_1,\dots,\vb{P}_N)=
    &\frac{1}{(2\uppi)^{3N/2}}
    \int\dd\vb{R}_1\cdots\dd\vb{R}_{N}\,\mathrm{e}^{-\mathrm{i}\sum_i\vb{P}_i\cdot\vb{R}_i}\notag\\
    &\times\Psi_X(\vb{R}_1,\dots,\vb{R}_N).
    \label{eq:tildepsix}
\end{align}
However, in the rotational-vibrational eigenstate, the position-space wavefunction is usually expressed as $\Psi_u$ in the generalized coordinate system $(u_a)\equiv(u_1,\dots,u_{3N})=(\vb{R}_\mathrm{G},\Theta,\vb{Q})$, where $\vb{R}_\mathrm{G}=(R_{\mathrm{G}X},R_{\mathrm{G}Y},R_{\mathrm{G}Z})=(u_1,u_2,u_3)$ are the center-of-mass coordinates, $\Theta$ are the Euler angles, and $\vb{Q}$ are the normal coordinates.\cite{wilson1955molecular, bunker2006moleculara}
For nonlinear molecules, $\Theta=(\phi,\theta,\chi)=(u_4,u_5,u_6)$ and $\vb{Q}=(Q_1,\dots,Q_{3N-6})=(u_7,\dots,u_{3N})$; for linear molecules, $\Theta=(\phi,\theta)=(u_4,u_5)$ and $\vb{Q}=(Q_1,\dots,Q_{3N-5})=(u_6,\dots,u_{3N})$.

Hence, to calculate $\Phi_X$ of \cref{eq:tildepsix}, the change of variables from $(X_a)$ to $(u_a)$ is necessary. 
According to the \emph{Podolsky trick},\cite{podolsky1928quantummechanically,wilson1955molecular,bunker2006moleculara} which is used in the derivation of the rotational-vibrational Hamiltonian for $\Psi_u$, the $\Psi_X$ is related to the $\Psi_u$ by 
\begin{align}
    \Psi_q=g_X^{-1/4}\Psi_X
    =g_u^{-1/4}\Psi_u,
    \label{eq:psix_to_psiq}
\end{align}
where $\Psi_q$ is the position-space wavefunction in the mass-weighted coordinate system $(q_a)\equiv(q_1,\dots,q_{3N})=(M_1^{1/2}\vb{R}_1,\dots,M_N^{1/2}\vb{R}_N)$ with the mass $M_i$ of the $i$th atom. 
$g_X$ and $g_u$ are the determinant of the matrices $(g_{X,ab})$ and $(g_{u,ab})$ whose components are expressed as
\begin{subequations}
\begin{align}
    &g_{X,ab}
    =\sum_c
    \frac{\partial q_c}{\partial X_a}
    \frac{\partial q_c}{\partial X_b},
    \label{eq:g_Xij}\\
    &g_{u,ab}
    =\sum_c
    \frac{\partial q_c}{\partial u_a}
    \frac{\partial q_c}{\partial u_b}.
    \label{eq:g_uij}
\end{align}    
\end{subequations}
The volume elements in the coordinates $(q_i)$, $(X_i)$, and $(u_i)$ are connected by $g_X$ and $g_u$ as follows:
\begin{align}
    \dd q_1\cdots\dd q_{3N}
    &=g_{X}^{1/2}\dd\vb{R}_1\cdots\dd\vb{R}_{N}=g_{u}^{1/2}\dd\vb{R}_\mathrm{G}\dd\Theta\dd\vb{Q}
    \label{eq:volume_element}
\end{align}
with $\dd\vb{R}_1\cdots\dd\vb{R}_{N}=\dd X_1\cdots\dd X_{3N}$ and $\dd\vb{R}_\mathrm{G}\dd\Theta\dd\vb{Q}=\dd u_1\cdots\dd u_{3N}$.
\cref{eq:psix_to_psiq} was originally introduced by Podolsky to satisfy \cref{eq:volume_element} and the normalization condition of the wavefunctions $\Psi_q$, $\Psi_X$, and $\Psi_u$,
\begin{align}
    \int\abs{\Psi_q}^2\dd q_1\cdots\dd q_{3N}
    &=\int\abs{\Psi_X}^2\dd\vb{R}_1\cdots\dd\vb{R}_{N}\notag\\
    &=\int\abs{\Psi_u}^2\dd\vb{R}_\mathrm{G}\dd\Theta\dd\vb{Q}=1.
\end{align}
From \cref{eq:psix_to_psiq,eq:volume_element}, we can rewrite $\Phi_X$ as

\begin{align}
    \Phi_X(\vb{P}_1,\dots,\vb{P}_N)
    =&\frac{1}{(2\uppi)^{3N/2}}
    \int\dd\vb{R}_\mathrm{G}\dd\Theta\dd\vb{Q}\,\mathrm{e}^{-\mathrm{i}\sum_i\vb{P}_i\cdot\vb{R}_i}\notag\\
    &\times\abs{|J|}^{1/2}\Psi_u(\vb{R}_\mathrm{G},\Theta,\vb{Q})
    ,
    \label{eq:tildepsix_from_X_to_q}
\end{align}
where $\abs{|J|}\equiv\sqrt{g_u/g_X}$ is the Jacobian of $(X_a)$ with respect to $(u_a)$.
\subsubsection{Separation of translational and internal motions}
\begin{figure}[ht]
  \includegraphics{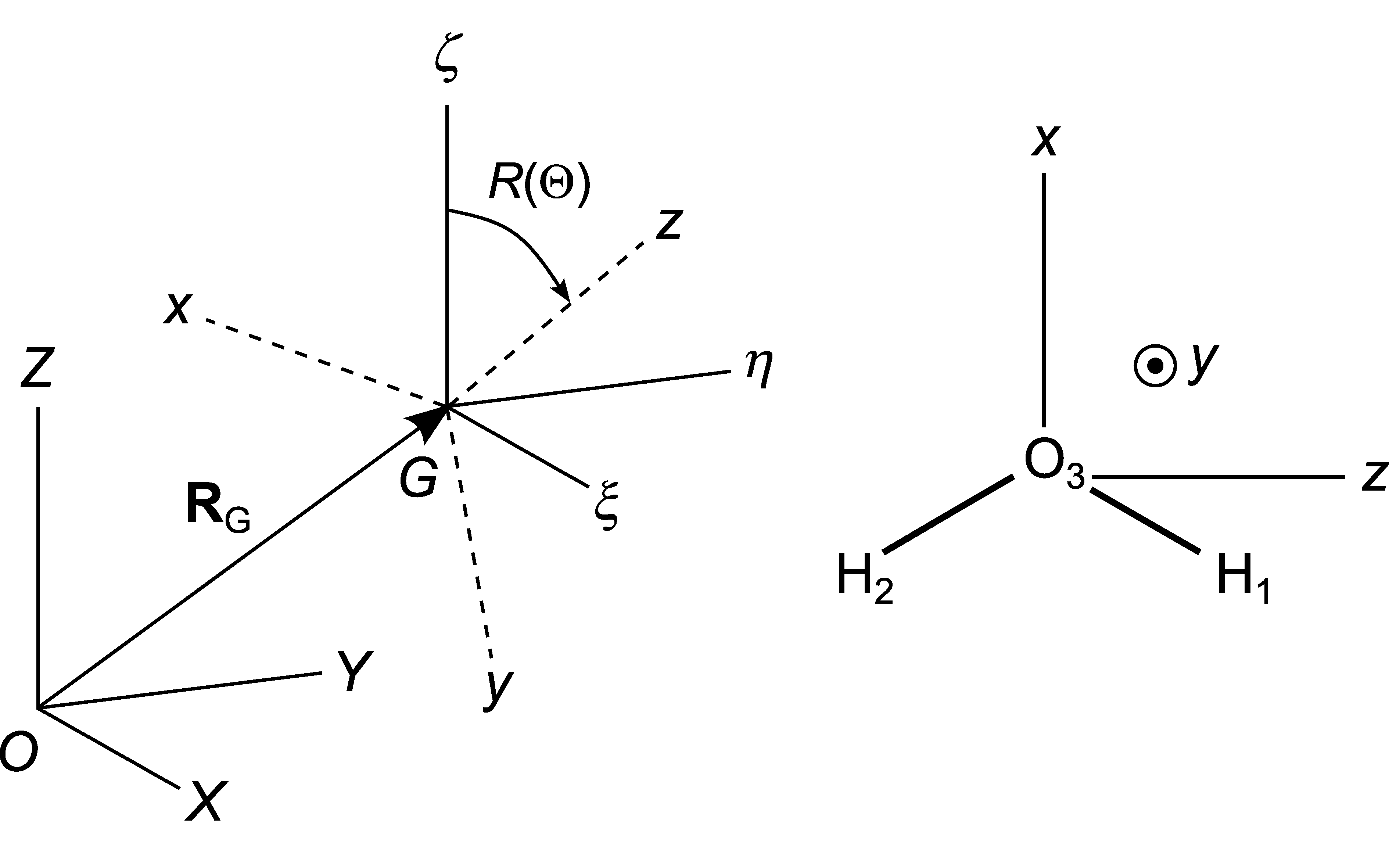}
  \caption{Spatial relationships of the space-fixed $(X,Y,Z)$, the center-of-mass $ (\xi,\eta,\zeta)$, and the molecule-fixed $(x,y,z)$ frames.
 }\label{fig:coordinates}
\end{figure}
In the absence of external fields, the position-space wavefunction $\Psi_u$ is written as the product of the translational $\Psi_u^\mathrm{T}$ and rotational-vibrational wavefunctions $\Psi_u^\mathrm{RV}$:\cite{wilson1955molecular, bunker2006moleculara}
\begin{align}
    \Psi_u(\vb{R}_\mathrm{G},\Theta,\vb{Q})
    =\Psi_u^\mathrm{T}(\vb{R}_\mathrm{G})\Psi_u^\mathrm{RV}(\Theta,\vb{Q}).
    \label{eq:psiq=psiqT*psiqRV}
\end{align}
Also, $\Phi_X$ can be separated into the two parts by setting the coordinate frames as shown in \cref{fig:coordinates}, as follows. 
Here the center-of-mass frame $ (\xi,\eta,\zeta)$ is moved by $\vb{R}_\mathrm{G}$ from the space-fixed frame $(X,Y,Z)$, and the molecule-fixed $(x, y, z)$ frame is converted from the $ (\xi,\eta,\zeta)$ frame by the rotation matrix $R(\Theta)$. 
Then, the coordinates $\vb{R}_i$ of the $i$th atom in the space-fixed frame $(X,Y,Z)$ are expressed as
\begin{align}
    \vb{R}_i
    =\vb{R}_\mathrm{G}+\boldsymbol{\uprho}_i
    \label{eq:rn}
\end{align}
with $\boldsymbol{\uprho}_i=(\rho_{i\xi},\rho_{i\eta},\rho_{i\zeta})$ being the coordinates in the $(\xi,\eta,\zeta)$ frame.
From the definition of the center-of-mass coordinates,
\begin{subequations}
\label{eq:MTrG&mirhoi}
    \begin{align}
    &\sum_iM_i\vb{R}_i=M_\mathrm{T}\vb{R}_\mathrm{G},\\
    &\sum_iM_i\boldsymbol{\uprho}_i=0,
\end{align}
\end{subequations}
where $M_\mathrm{T}=\sum_i M_i$ is the total mass of the molecule. On the other hand, the momentum $\vb{P}_i$ of the $i$th atom is expressed as 
\begin{align}
    \vb{P}_i=\frac{M_i}{M_\mathrm{T}}\vb{P}_\mathrm{T}+\boldsymbol{\uppi}_i
    \label{eq:Pn}
\end{align}
in terms of the total momentum $\vb{P}_\mathrm{T}=(P_{\mathrm{T}X},P_{\mathrm{T}Y},P_{\mathrm{T}Z})$ and the momentum $\boldsymbol{\uppi}_i=(\pi_{i\xi},\pi_{i\eta},\pi_{i\zeta})$ of the $i$th atom in the center-of-mass frame, which satisfy
\begin{subequations}
\label{eq:PTandpi}
\begin{align}
    &\sum_i\vb{P}_i=\vb{P}_\mathrm{T},\\
    &\sum_i\boldsymbol{\uppi}_i=0.
    \label{eq:sum_pi=0}
\end{align}    
\end{subequations}
Using \cref{eq:rn,eq:MTrG&mirhoi,eq:Pn,eq:PTandpi}, we can write the phase of the plane wave in \cref{eq:tildepsix_from_X_to_q} as 
\begin{align}
    \sum_i\vb{P}_i\cdot\vb{R}_i
    =\vb{P}_\mathrm{T}\cdot\vb{R}_\mathrm{G}
    +\sum_i\boldsymbol{\uppi}_i\cdot\boldsymbol{\uprho}_i.
    \label{eq:pr=ptrg+pirho}
\end{align}
Substituting \cref{eq:psiq=psiqT*psiqRV,eq:pr=ptrg+pirho} into \cref{eq:tildepsix_from_X_to_q}, we obtain
\begin{align}
    \Phi_X(\vb{P}_1,\dots,\vb{P}_N)
    =\Phi_X^\mathrm{T}(\vb{P}_\mathrm{T})
    \Phi_X^\mathrm{RV}(\boldsymbol{\uppi}_1,\dots,\boldsymbol{\uppi}_N),
    \label{eq:tildepsix=tildepsixT*tildepsixRV}
\end{align}
where 
\begin{align}
    &\Phi_X^\mathrm{T}(\vb{P}_\mathrm{T})
    =\frac{1}{(2\uppi)^{3/2}}
    \int\Psi_u^\mathrm{T}(\vb{R}_\mathrm{G})
    \mathrm{e}^{-\mathrm{i}\vb{P}_\mathrm{T}\cdot\vb{R}_\mathrm{G}}
    \dd\vb{R}_\mathrm{G},\\
    &\Phi_X^\mathrm{RV}(\boldsymbol{\uppi}_1,\dots,\boldsymbol{\uppi}_N)
    =\frac{1}{(2\uppi)^{(3N-3)/2}}\int\dd\Theta\dd\vb{Q}\,\mathrm{e}^{-\mathrm{i}\sum_i\boldsymbol{\uppi}_i\cdot\boldsymbol{\uprho}_i}\notag\\
    &\hphantom{\Phi_X^\mathrm{RV}(\boldsymbol{\uppi}_1,\dots,\boldsymbol{\uppi}_N)={}}\times\abs{|J|}^{1/2}\Psi_u^\mathrm{RV}(\Theta,\vb{Q})
    \label{eq:tildepsixRV}.
\end{align}
In the derivation of \cref{eq:tildepsix=tildepsixT*tildepsixRV}, we have made use of the fact that $\abs{|J|}$ is independent of $\vb{R}_\mathrm{G}$, as shown in the \cref{appendix}.

From \cref{eq:PTandpi,eq:tildepsix=tildepsixT*tildepsixRV}, we can rewrite the atomic momentum distribution as
\begin{align}
    n_s(\vb{P}_s)=
    \int n^\mathrm{T}(\vb{P}_\mathrm{T})
    n^\mathrm{RV}_s\left(\vb{P}_s-\frac{M_s}{M_\mathrm{T}}\vb{P}_\mathrm{T}\right)
    \dd\vb{P}_\mathrm{T},
    \label{eq:nn=nT*nRV}
\end{align}
where $n^\mathrm{T}=\abs{\Phi_X^\mathrm{T}}^2$ and
\begin{align}
    n^\mathrm{RV}_s(\boldsymbol{\uppi}_s)
    =&\int_U\dd\boldsymbol{\uppi}_1\cdots\dd\boldsymbol{\uppi}_{s-1}\dd\boldsymbol{\uppi}_{s+1}\cdots\dd\boldsymbol{\uppi}_{N}\notag\\
    &\times\abs{\Phi_X^\mathrm{RV}
    \left(\boldsymbol{\uppi}_1,\dots,\boldsymbol{\uppi}_N\right)}^2
    \label{eq:nRV_pi}
\end{align}
with $\boldsymbol{\uppi}_s=\vb{P}_s-M_s\vb{P}_\mathrm{T}/M_\mathrm{T}$ and the momentum space $U$ which satisfies $\sum_i\boldsymbol{\uppi}_i=0$.
\cref{eq:nn=nT*nRV} means that we can calculate $n_s$ by the convolution of the atomic momentum distribution $n_s^\mathrm{RV}$ due to the internal (rotational and vibrational) motion and the translational momentum distribution $n^\mathrm{T}$. The energy eigenstates of translational motion in free molecular systems are momentum eigenstates, and hence it is not necessary to consider the translational contribution as long as stationary states are considered.
Hereafter, we will focus only on the internal motion by setting $n^\mathrm{T}=\delta(\vb{P}_\mathrm{T})$ so that the atomic momentum distribution is expressed as
\begin{align}
    n_s(\vb{P}_s)=n_s^\mathrm{RV}(\vb{P}_s).
    \label{eq:nn=nnRV}
\end{align}
\subsubsection{The coordinates in the center-of-mass frame and the Jacobian expressed by the generalized coordinates}
\label{subsec:Expressions for the center-of-mass coordinates and the Jacobian}
To obtain $n_s^\mathrm{RV}$ for a given $\Psi_u^\mathrm{RV}$, we have to evaluate the integral in \cref{eq:nRV_pi} using $\Phi_X^\mathrm{RV}$ of \cref{eq:tildepsixRV}. In this case, it is necessary to express both the coordinates $\boldsymbol{\uprho}_i$ in the center-of-mass frame and the Jacobian $\abs{|J|}$ as functions of the generalized coordinates. 

Since the molecule-fixed frame $(x,y,z)$ is converted from the center-of-mass frame by the rotation matrix $R(\Theta)$ as shown in \cref{fig:coordinates}, the coordinates $\vb{r}_i=(r_{ix},r_{iy},r_{iz})$ in the $(x,y,z)$ frame and $\boldsymbol{\uprho}_i$ satisfy the following relationship:
\begin{align}
    \boldsymbol{\uprho}_i
    =R(\Theta)\vb{r}_i,
    \label{eq:rho=Rr}
\end{align}
where we treat $\boldsymbol{\uprho}_i$ and $\vb{r}_i$ as column vectors.
In the Z-Y-Z convention of the Euler angles, $R(\Theta)$ is expressed as
\begin{align}
    &R(\Theta)=\begin{pmatrix}
    \mathrm{c}\theta\mathrm{c}\phi\mathrm{c}\chi-\mathrm{s}\phi\mathrm{s}\chi&
    -\mathrm{c}\theta \mathrm{c}\phi \mathrm{s}\chi-\mathrm{s}\phi \mathrm{c}\chi&
    \mathrm{s}\theta \mathrm{c}\phi\\
    \mathrm{c}\theta \mathrm{s}\phi \mathrm{c}\chi+\mathrm{c}\phi \mathrm{s}\chi&
    -\mathrm{c}\theta \mathrm{s}\phi \mathrm{s}\chi+\mathrm{c}\phi \mathrm{c}\chi&
    \mathrm{s}\theta \mathrm{s}\phi\\
    -\mathrm{s}\theta \mathrm{c}\chi&\mathrm{s}\theta \mathrm{s}\chi&\mathrm{c}\theta
    \end{pmatrix},
    \label{eq:Rotation_matrix}
\end{align}
where c and s represent cosine and sine, respectively. For linear molecules, $\chi$ is an arbitrary function of $(\phi,\theta)$,\cite{watson1970vibrationrotation} and we set $\chi=0$.
According to the theory of molecular vibration, 
we can separate $\vb{r}_i$ into the equilibrium coordinates $\vb{r}_i^0=(r_{ix}^0,r_{iy}^0,r_{iz}^0)$ and the displacements $\Delta\vb{r}_i=(\Delta r_{ix},\Delta r_{iy},\Delta r_{iz})$ which are further decomposed into $N_\mathrm{V}$ normal modes:\cite{wilson1955molecular}
\begin{subequations}
\label{eq:r=r0+dr}
\begin{align}
    &\vb{r}_i=\vb{r}_i^0+\Delta\vb{r}_i,
    \\
    &\Delta\vb{r}_i(\vb{Q})
    =M_i^{-1/2}\sum_{k=1}^{N_\mathrm{V}}
    \vb{l}_{i,k}Q_k
\end{align}    
\end{subequations}
with the vibrational degrees of freedom $N_\mathrm{V}=3N-6\text{ (or }3N-5)$.
The expansion coefficients $\vb{l}_{i,k}=(l_{i,k}^x,l_{i,k}^y,l_{i,k}^z)$ satisfy the orthogonality and the Eckart conditions:\cite{watson1968simplification,eckart1935studies}
\begin{subequations}
    \label{eq:condition_of_l}
    \begin{align}
    &\sum_i\vb{l}_{i,k}\cdot\vb{l}_{i,l}=\delta_{kl},
    \label{eq:orthogonality_of_l}\\
    &\sum_iM_i^{1/2}\vb{l}_{i,k}=0,\\
    &\sum_iM_i^{1/2}\left(\vb{r}_i^0\times\vb{l}_{i,k}\right)=0.
    \label{eq:Eckart_condition_of_l}
\end{align}
\end{subequations}
Using 
\cref{eq:rho=Rr,eq:r=r0+dr}, we can rewrite $\boldsymbol{\uprho}_i$ as
\begin{align}
    \boldsymbol{\uprho}_i
    =R(\Theta)
    \left(\vb{r}_i^0
    +M_i^{-1/2}\sum_{k}\vb{l}_{i,k}Q_k\right),
    \label{eq:rhon=R(rho'n0+drho_n)}
\end{align}
which enables us to calculate the Jacobian $\abs{|J|}$.
After some mathematical transformations (see \cref{appendix}), we obtain
\begin{align}
    \abs{|J|}=(g_u/g_X)^{1/2}=G^{1/2}\sin\theta.
    \label{eq:absJ_expressions}
\end{align}
The general form of $G$ for nonlinear molecules is given by
\begin{align}
    G(\vb{Q})=
    M_\mathrm{T}^3\det(I'_{\alpha\beta})/\prod_iM_i^3,
\end{align}
where $I'_{\alpha\beta}$ is the component of the effective inertia tensor (with respect to the molecule-fixed frame) defined by\cite{watson1968simplification}
\begin{align}
    I'_{\alpha\beta}(\vb{Q})=I_{\alpha\beta}(\vb{Q})-\sum_{klm}\zeta^\alpha_{km}\zeta^\beta_{lm}Q_kQ_l
\end{align}
with
\begin{align}
    &I_{\alpha\beta}(\vb{Q})=\sum_{i}M_i(\delta_{\alpha\beta}r^2_i-r_{\alpha i}r_{\beta i}),\\
    &\zeta^\alpha_{km}=\sum_{\beta\gamma}\varepsilon_{\alpha\beta\gamma}\sum_{i}l_{i,k}^\beta l_{i,m}^\gamma.
\end{align}
$I_{\alpha\beta}$ is the instantaneous inertia tensor, and $\zeta^\alpha_{km}$ is the Coriolis coupling coefficient.
For linear molecules, $G$ is written as
\begin{align}
    G(\vb{Q})
    =M_\mathrm{T}^3I'^2/\prod_i M_i^3.
\end{align}
The effective moment of inertia $I'$ is related to $I'_{\alpha\beta}$ as follows:\cite{watson1970vibrationrotation,amat1958coefficients}
\begin{subequations}
\label{eq:inertia_tensor&momentum}
\begin{align}
    &I'_{\alpha\beta}(\vb{Q})
    =\epsilon_{\alpha\beta}I'(\vb{Q}),
    \\
    &\epsilon_{\alpha\beta}=\delta_{\alpha\beta}-\delta_{\alpha z}\delta_{\beta z},
\end{align}
\end{subequations}
where the $z$-axis is the molecular axis of linear molecules.
\subsection{Rigid-rotor-harmonic-oscillator approximation}
Since $\boldsymbol{\uprho}_i$ and $\abs{|J|}$ are obtained as the functions of $(\Theta,\vb{Q})$, we can calculate $\Phi_X^\mathrm{RV}$ by using \cref{eq:tildepsixRV}. 
However, to obtain $\Phi_X^\mathrm{RV}$, the $(3N-3)$-dimensional integral must be performed for each point of $(\boldsymbol{\uppi}_1,\dots,\boldsymbol{\uppi}_N)$ in the momentum space. 
In order to reduce the computational cost, we introduce the rigid-rotor-harmonic-oscillator (RRHO) approximation.

Using \cref{eq:rhon=R(rho'n0+drho_n)}, we can write the phase of plane wave in \cref{eq:tildepsixRV} as 
\begin{align}
    \sum_i\boldsymbol{\uppi}_i\cdot\boldsymbol{\uprho}_i
    =\sum_i\boldsymbol{\uppi}_i\cdot\boldsymbol{\uprho}_i^0
    +\sum_k P_{Q_k} Q_k,
    \label{eq:pirho=pirho0+PQ}
\end{align}
where
\begin{subequations}
\begin{align}
    &\boldsymbol{\uprho}_i^0(\Theta)
    =R(\Theta)\vb{r}_i^{0},\\
    &P_{Q_k}(\Theta)
    =\sum_i\boldsymbol{\uppi}_i^\top R(\Theta)M_i^{-1/2}\vb{l}_{i,k}.
\end{align}
\end{subequations}
Substituting \cref{eq:pirho=pirho0+PQ,eq:absJ_expressions} into \cref{eq:tildepsixRV}, we rewrite
\begin{align}
    &\Phi_X^\mathrm{RV}(\boldsymbol{\uppi}_1,\dots,\boldsymbol{\uppi}_N)
    =\frac{1}{(2\uppi)^{N_\mathrm{R}/2}}
    \int\sin^{1/2}\theta\dd\Theta\,
    \mathrm{e}^{-\mathrm{i}\sum_i\boldsymbol{\uppi}_i\cdot\boldsymbol{\uprho}_i^0}
    \notag\\
    &\hphantom{\Phi^\mathrm{RV}(\boldsymbol{\uppi}}\times\left[
    \frac{1}{(2\uppi)^{N_\mathrm{V}/2}}
    \int G^{1/4}\Psi_u^\mathrm{RV}
    \mathrm{e}^{-\mathrm{i}\sum_k P_{Q_k} Q_k}\dd\vb{Q}\right]
    \label{eq:rewriting_PhiXRV}
\end{align}
with $N_\mathrm{R}=3\text{ (or 2)}$ being the rotational degrees of freedom.
The integration with respect to $\vb{Q}$ in \cref{eq:rewriting_PhiXRV} is regarded as the Fourier transform of $\Psi_u^\mathrm{RV}$. 
By applying the RRHO approximation to $\Psi_u^\mathrm{RV}$ and $G$, we will calculate this Fourier transform analytically.

In the RRHO approximation, $\Psi_u^\mathrm{RV}$ is separated into the rigid rotor and harmonic oscillator wavefunctions:\cite{wilson1955molecular,bunker2006moleculara}
\begin{align}
    \Psi_u^\mathrm{RV}(\Theta,\vb{Q})\simeq
    \left[\Psi_u^\mathrm{R}(\Theta)\sin^{1/2}\theta\right]\Psi_u^\mathrm{V}(\vb{Q}).
    \label{eq:PsiqRV=PsiqR*PsiqV}
\end{align}
The normalization conditions of $\Psi_u^\mathrm{R}$ and $\Psi_u^\mathrm{V}$ are given by
\begin{align}
    \int\abs{\Psi_u^\mathrm{R}}^2
    \sin\theta\dd\Theta=\int\abs{\Psi_u^\mathrm{V}}^2
    \dd\vb{Q}=1
\end{align}
with $\dd\Theta=\dd\phi\dd\theta\dd\chi\text{ (or } \dd\phi\dd\theta)$.
$\Psi_u^\mathrm{V}$ is expressed as the product of the harmonic oscillator wavefunctions for each normal mode of the molecule:
\begin{align}
    \Psi_u^\mathrm{V}(\vb{Q})=&\prod_{k}\psi_{v_k}({Q_k})
    \label{eq:PsiqV}
\end{align}
with
\begin{align}
    \psi_{v_k}({Q_k})=&\left(\frac{1}{2^{v_k}v_k!}
    \sqrt{\frac{\omega_k}{\uppi\hbar}}\right)^{1/2}\notag\\
    &\times\exp(-\frac{\omega_k{Q}_k^2}{2\hbar})H_{v_k}\left(\sqrt{\frac{\omega_k}{\hbar}}{Q}_k\right),
\end{align}
where $v_k$ is the vibrational quantum number, $\omega_k$ is the angular frequency, and $H_{v_k}$ is the $v_k$th-order Hermite polynomial.
In the rigid rotor approximation of $G$, we write
\begin{align}
    G(\vb{Q})
    \simeq G(0)\equiv G_0.
    \label{eq:abs|J|=C0sin_th}
\end{align}
This treatment is equivalent to $I'_{\alpha\beta}\simeq I_{\alpha\beta}(0)\equiv I_{\alpha\beta}^0$ used in the RRHO approximation of the rotation-vibration Hamiltonian.\cite{wilson1955molecular,bunker2006moleculara}
Substituting \cref{eq:PsiqRV=PsiqR*PsiqV,eq:abs|J|=C0sin_th} into \cref{eq:rewriting_PhiXRV}, we obtain
\begin{align}
    \Phi_X^\mathrm{RV}(\boldsymbol{\uppi}_1,\dots,\boldsymbol{\uppi}_N)
    =&\frac{G_0^{1/4}}{(2\uppi)^{N_\mathrm{R}/2}}
    \int\sin\theta\dd\Theta\notag\,\mathrm{e}^{-\mathrm{i}\sum_i\boldsymbol{\uppi}_i\cdot\boldsymbol{\uprho}_i^0}\\
    &\times\Psi_u^\mathrm{R}(\Theta)
    \Phi_u^\mathrm{V}(\vb{P}_Q)
    \label{eq:PhiXRV_approximation}
\end{align}
with $\vb{P}_Q(\Theta)=(P_{Q_1},\dots,P_{Q_{N_\mathrm{V}}})$.
$\Phi_u^\mathrm{V}$ is the Fourier transform of $\Psi_u^\mathrm{V}$ and is written as
\begin{align}
    \Phi_u^\mathrm{V}(\vb{P}_Q)
    &=\prod_{k}\phi_{v_k}(P_{Q_k}),
\end{align}
where
\begin{align}
    \phi_{v_k}(P_{Q_k})=&
    (-\mathrm{i})^{v_k}
    \left(\frac{1}{2^{v_k}v_k!}
    \sqrt{\frac{\hbar}{\uppi\omega_k}}\right)^{1/2}\notag\\
    &\times\exp(-\frac{\hbar P^2_{Q_k}}{2\omega_k})
    H_{v_k}\left(\sqrt{\frac{\hbar}{\omega_k}}P_{Q_k}\right).
\end{align}
Finally, we can calculate the atomic momentum distribution within the RRHO framework by using \cref{eq:nRV_pi,eq:nn=nnRV,eq:PhiXRV_approximation}.

\section{Computational details}
\label{Computational Details}
To investigate the quantum effects on the atomic momentum distribution, the calculations of atomic momentum distributions were performed within the two kinds of frameworks, i.e., the quantum and semiclassical ones, dealing with molecular free rotations in quantum and classical ways, respectively.
In both cases, we set $n^\mathrm{T}=\delta(\vb{P}_\mathrm{T})$ and utilize the RRHO approximation. 
The target are nonlinear and linear triatomic molecules, \ce{H2O} and \ce{CO2}, in their rotational-vibrational ground state ($T=\SI{0}{K}$).

\subsection{Quantum framework}
In the rotational-vibrational ground state (the angular momentum quantum number $J=0$ and $\{v_k\}=0$), the rotational wavefunction of nonlinear (or linear) molecules is given by
\begin{align}
    \Psi_u^\mathrm{R}(\Theta)
    =\frac{1}{\sqrt{8\uppi^2}}
    \left(\text{or }\frac{1}{\sqrt{4\uppi}}
    \right).
    \label{eq:PsiqR_groundstate}
\end{align}
Putting \cref{eq:PsiqR_groundstate} into \cref{eq:PhiXRV_approximation}, it is evident that $\Phi_X^\mathrm{RV}$ has the following rotational symmetry:
\begin{align}
    \Phi_X^\mathrm{RV}(R\boldsymbol{\uppi}_1,\dots,R\boldsymbol{\uppi}_N)
    =\Phi_X^\mathrm{RV}
    (\boldsymbol{\uppi}_1,\dots,\boldsymbol{\uppi}_N).
    \label{eq:rotational_symmetry_of_PhiX_}
\end{align}
Therefore, the atomic momentum distribution of \cref{eq:nRV_pi} becomes isotropic in the space-fixed frame:
\begin{align}
    n_s^\mathrm{RV}(\boldsymbol{\uppi}_s)
    =n_s^\mathrm{RV}(\pi_s)
    \label{eq:isotropic nnRV}.
\end{align}
From \cref{eq:nn=nT*nRV,eq:nn=nnRV,eq:isotropic nnRV,eq:nRV_pi}, the atomic momentum distribution of the target atom A in a triatomic molecule (ABC) can be written as
\begin{align}
    n_\mathrm{A}(P_\mathrm{A})=
    \int\abs{\Phi_X^\mathrm{RV}
    (\vb{P}_\mathrm{A},\vb{P}_\mathrm{B},
    -\vb{P}_\mathrm{A}-\vb{P}_\mathrm{B})}^2
    \dd\vb{P}_\mathrm{B},
    \label{eq:n1=intpsiRV}
\end{align}
\begin{subequations}
where we replace $\boldsymbol{\uppi}_\mathrm{B}$ with $\vb{P}_\mathrm{B}$. Since the atomic momentum distribution in the ground state is isotropic, we just consider one-dimensional distribution for $\vb{P}_\mathrm{A}$ and express $\vb{P}_\mathrm{B}$ by using the cylindrical coordinates, 
\begin{align}
    &\vb{P}_\mathrm{A}=(0,0,P_\mathrm{A}),\\
    &\vb{P}_\mathrm{B}=(P_{\mathrm{B}r}\cos\varphi_\mathrm{B},P_{\mathrm{B}r}\sin\varphi_\mathrm{B},P_{\mathrm{B}Z}).
\end{align}    
\end{subequations}
Using  \cref{eq:rotational_symmetry_of_PhiX_} with rotations $R_Z$ around the $Z$-axis, $\Phi_X^\mathrm{RV}$ has cylindrical symmetry about $\vb{P}_\mathrm{B}$:
\begin{align}
    \Phi_X^\mathrm{RV}
    (\vb{P}_\mathrm{A},R_Z\vb{P}_\mathrm{B})
    =\Phi_X^\mathrm{RV}
    (\vb{P}_\mathrm{A},\vb{P}_\mathrm{B}).
\end{align}
Therefore, we can rewrite \cref{eq:n1=intpsiRV} as
\begin{align}
    n_\mathrm{A}(P_\mathrm{A})
    =2\uppi\int\abs{\Phi_X^\mathrm{RV}
    (P_\mathrm{A},P_{\mathrm{B}r},P_{\mathrm{B}Z})}^2
    P_{\mathrm{B}r}\dd P_{\mathrm{B}r}\dd P_{\mathrm{B}Z}
    \label{eq:nP_cylindrical}
\end{align}
with $\varphi_\mathrm{B}=0$.

In order to calculate $\Phi_X^\mathrm{RV}$ the physical quantities $\vb{r}_i^0,\vb{l}_{i,k}\text{, and }\omega_k$ for \ce{H2O} and \ce{CO2} were calculated at the HF/6-31G(d,p) level using the Gaussian16 program.\cite{g16} In determining $\vb{r}_i^0$ and $\vb{l}_{i,k}$, we set the axes of the molecule-fixed frame as $(x,y,z)=(b,c,a)$ with the principal axes $(a,b,c)$ at the equilibrium molecular structure, where the principal moments of inertia satisfy $I_{aa}^0<I_{bb}^0<I_{cc}^0$.
For \ce{CO2}, the $a$ axis coincides with the molecular axis, i.e., the $z$ axis.

In the numerical evaluation of \cref{eq:PhiXRV_approximation}, the 
integrand is periodic for $\phi$ and $\chi$, but not for $\theta$. Therefore, we used the trapezoidal integration for $\phi$ and $\chi$, and the tanh-sinh quadrature for $\theta$ with
\begin{align}
    \cos\theta=\tanh(\frac{\uppi}{2}\sinh t),
\end{align}
where we set $\abs{t}\leq3$.
For \ce{H2O}, the numbers of sampled points are 100, 200, and 100 for $\phi$, $t$, and $\chi$, respectively; for \ce{CO2}, 100 and 400 for $\phi$ and $t$, respectively. 
\begin{table}[ht]
    \centering
    \caption{The step size $\Delta P$ and the ranges of $P_\mathrm{A}$, $P_{\mathrm{B}r}$, and $P_{\mathrm{B}Z}$.
    The unit $a_0$ is the Bohr radius and  $1\,a_0^{-1}=1.89\,\si{\angstrom^{-1}}$.}
    \begin{ruledtabular}
    \begin{tabular}{ccccc}
         Molecule(Atom A,B)
         &$\Delta P/a_0^{-1}$
         &$P_\mathrm{A}/a_0^{-1}$
         &$P_{\mathrm{B}r}/a_0^{-1}$
         &$P_{\mathrm{B}Z}/a_0^{-1}$\\ \hline
         \ce{H2O}(\ce{H1},\ce{H2})&0.1&0--20&0--15&$-15$--15\\
         \ce{CO2}(\ce{O1},\ce{O2})&0.1&0--40&0--15&$-50$--30\\
    \end{tabular}        
    \end{ruledtabular}
    \label{tab:integral interval}
\end{table}

From $\Phi_X^\mathrm{RV}$ obtained in the way described above, we calculated $n_\mathrm{A}$ of \cref{eq:nP_cylindrical} by using the trapezoidal integration. 
The step size $\Delta P$ and the ranges of the momentum $P_\mathrm{A}$, $P_{\mathrm{B}r}$, and $P_{\mathrm{B}Z}$ are shown in \cref{tab:integral interval}. 
To evaluate the accuracy of these integration, we calculated the squared norm $\abs{\abs{\Phi}}^2$ and the average kinetic energy $\ev{E_{\mathrm{k}}}$, which are obtained in the case of isotropic momentum distribution as
\begin{align}
    &\abs{\abs{\Phi}}^2=\braket{\Phi}{\Phi}=\int_0^\infty4\uppi n_\mathrm{A}(P_\mathrm{A})P_\mathrm{A}^2\dd P_\mathrm{A},\\
    &\ev{{E}_{\mathrm{k}}}=\ev{\frac{P_\mathrm{A}^2}{2M_\mathrm{A}}}=\frac{1}{2M_\mathrm{A}}\int_0^\infty4\uppi n_\mathrm{A}(P_\mathrm{A})P_\mathrm{A}^4\dd P_\mathrm{A}.\label{eq:FQ_Ekin}
\end{align}

\subsection{Semiclassical framework}
The semiclassical theory, which treats the molecular rotation as classical motion while the vibration as quantum, has been already well developed,\cite{colognesi2001deep,ivanov1967interaction} and we can calculate $n_s$ and $\ev{E_{\mathrm{k}}}$ at $T=\SI{0}{K}$ by using
\begin{align}
    &n_s(P_s)=
    \frac{1}{4\uppi}\int\frac{1}{\sqrt{8\uppi^3}\sigma^3_s(\vu{p})}
    \exp(-\frac{P^2_s}{2\sigma^2_s(\vu{p})})
    \dd\vu{p},\\
    &\ev{E_{\mathrm{k}}}
    =\frac{\hbar^2}{2M_s}\left[\sigma^2_s(\vu{x})
    +\sigma^2_s(\vu{y})
    +\sigma^2_s(\vu{z})\right],
    \label{eq:SC_Ekin}
\end{align}
where
\begin{subequations}
\begin{align}
    &\hbar^2\sigma_s^2(\vu{p})
    =M_s\sum_k
    \frac{\hbar\omega_k}{2}
    \left(\vb{l}_{s,k}\cdot\vu{p}\right)^2,\\
    &\vu{p}=(\sin\theta_p\cos\phi_p,\sin\theta_p\sin\phi_p,\cos\theta_p).
\end{align}    
\end{subequations}

\section{Results and Discussion}
\label{Results and Discussion}
\begin{figure}[ht]
  \includegraphics{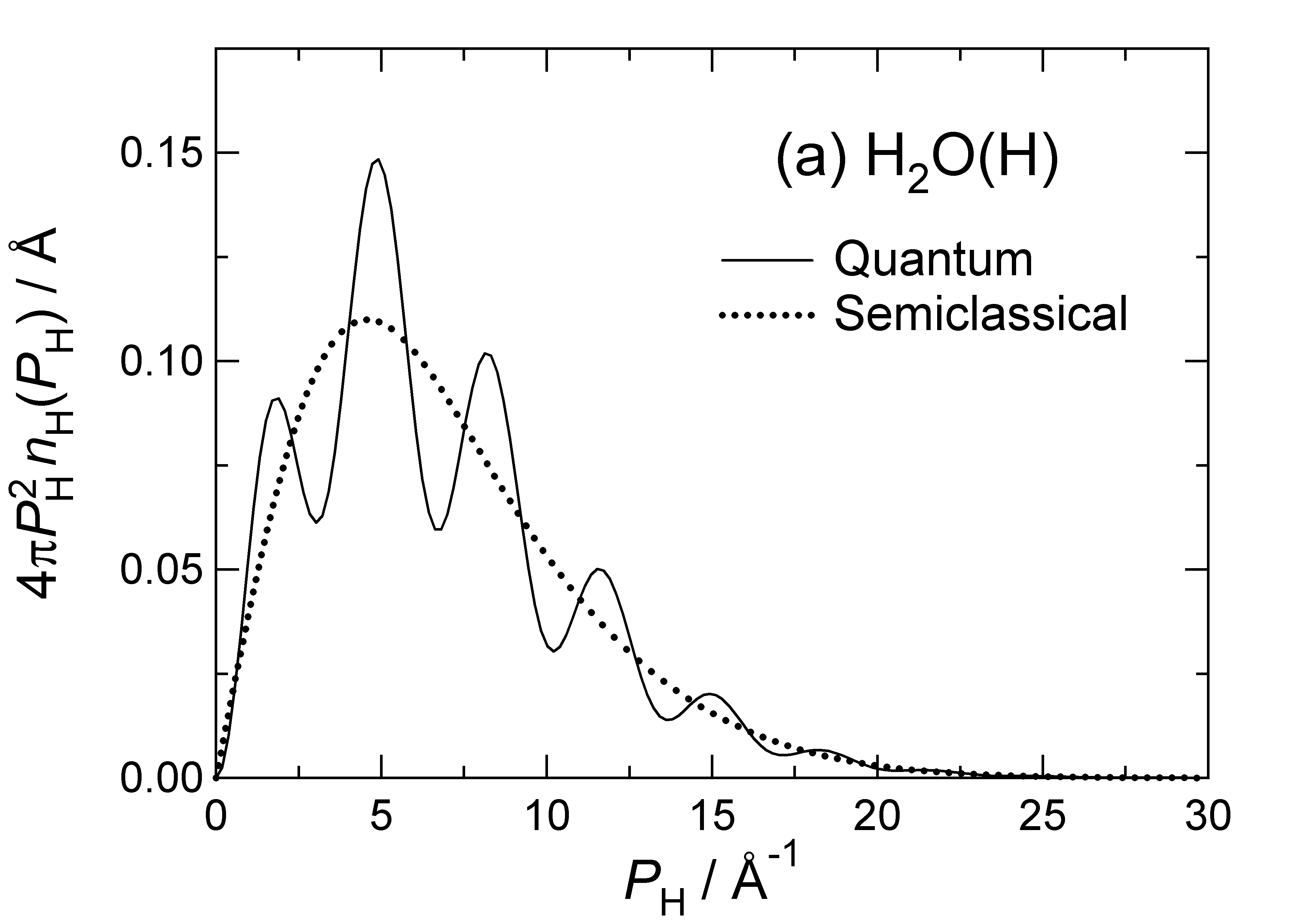}
  \includegraphics{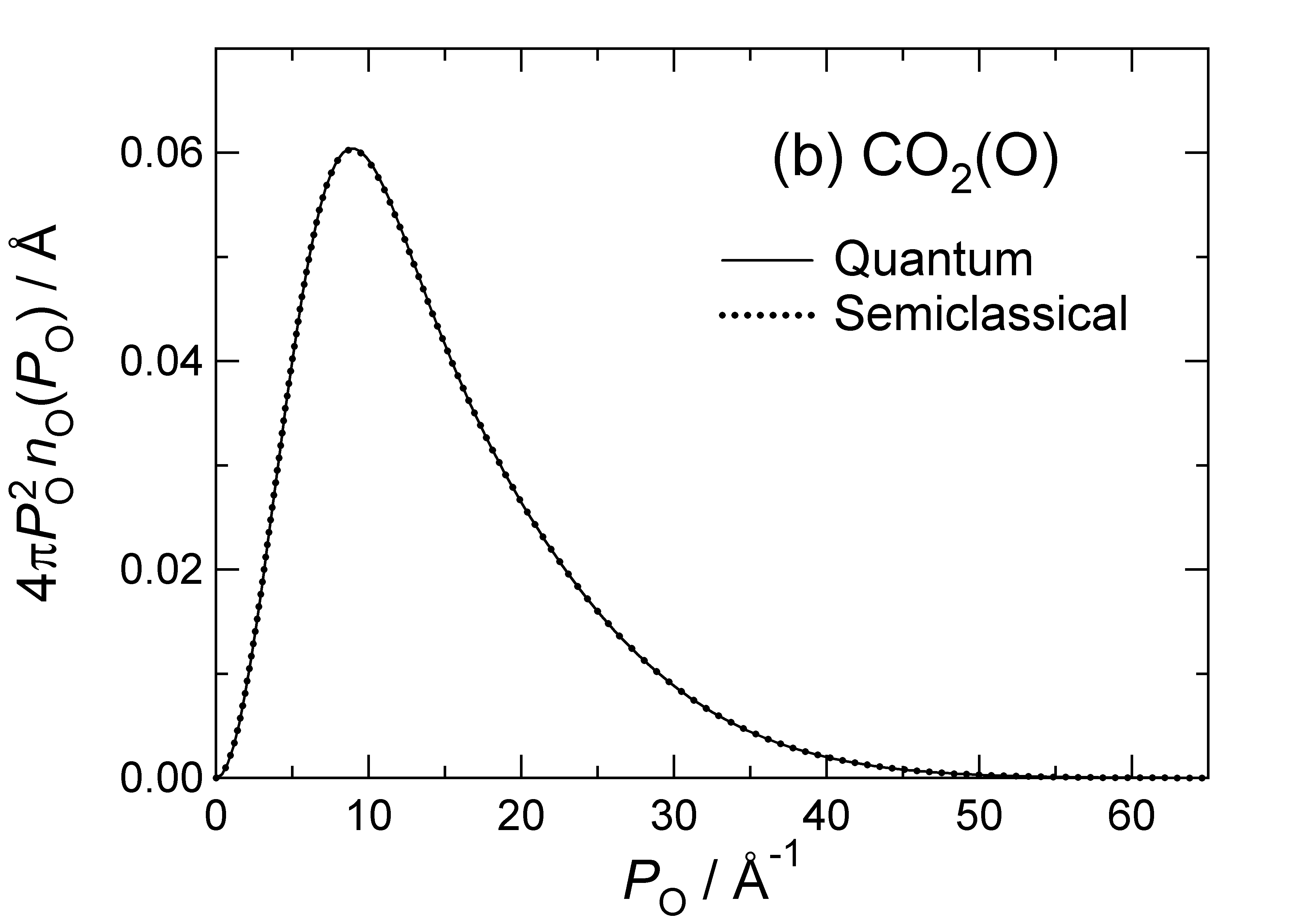}
  \caption{The atomic momentum distributions $4\uppi P_s^2\,n_s(P_s)$ in the ground state:
  (a)\,H atom in \ce{H2O} and (b)\,O atom in \ce{CO2}. 
  The solid and dotted lines represent the results of the quantum and semiclassical $(T=\SI{0}{K})$ calculations, respectively.}\label{fig:p^2np}
\end{figure}

\cref{fig:p^2np} shows the calculated momentum distributions of the \ce{H1} atom in \ce{H2O} and those of the \ce{O1} atom in \ce{CO2}, which were multiplied by a factor of $4{\pi}P^2_s$ in order to highlight structures appeared in the distributions. The calculated values for the squared norm $\abs{\abs{\Phi}}^2$ are listed in \cref{tab:innerproduct and ave. kinetic energy}. 
They are almost unity, which demonstrates the validity of the RRHO approximation. In addition, the $\ev{E_\mathrm{k}}$ values calculated by the quantum treatment with \cref{eq:FQ_Ekin} are in reasonable agreement with the theoretical values estimated by the semiclassical treatment with \cref{eq:SC_Ekin}. 
From \cref{fig:p^2np}, it is evident that an oscillatory structure appears in the H-atom momentum distribution in \ce{H2O} calculated within the framework of the quantum theory. On average, the result of the quantum calculation is in good agreement with that of the semiclassical calculation. On the other hand, no oscillatory behavior is found in the quantum result for the O-atom momentum distribution in \ce{CO2}.

\begin{table}[ht]
    \centering
    \caption{The calculated and theoretical values of the squared norm $\abs{\abs{\Phi}}^2$ and the mean kinetic energy $\ev{E_\mathrm{k}}$.
    The theoretical values of $\ev{E_\mathrm{k}}$ were estimated by \cref{eq:SC_Ekin}.}
    \begin{ruledtabular}
    \begin{tabular}{cccccc}
         \multirow{1}[3]{*}
         {Target Atom in Molecule}
         &\multicolumn{2}{c}{$\abs{\abs{\Phi}}^2$}
         &\multicolumn{2}{c}{$\ev{E_\mathrm{k}}/\si{meV}$}\\
         &Calc.
         &Theo.
         &Calc.
         &Theo.\\\hline
         \ce{H in H2O}&1.03&1&132&134\\
         \ce{O in CO2}&0.99&1&37.1&36.8\\
    \end{tabular}        
    \end{ruledtabular}
    \label{tab:innerproduct and ave. kinetic energy}
\end{table}

\begin{figure}[ht]
  \centering
  \includegraphics{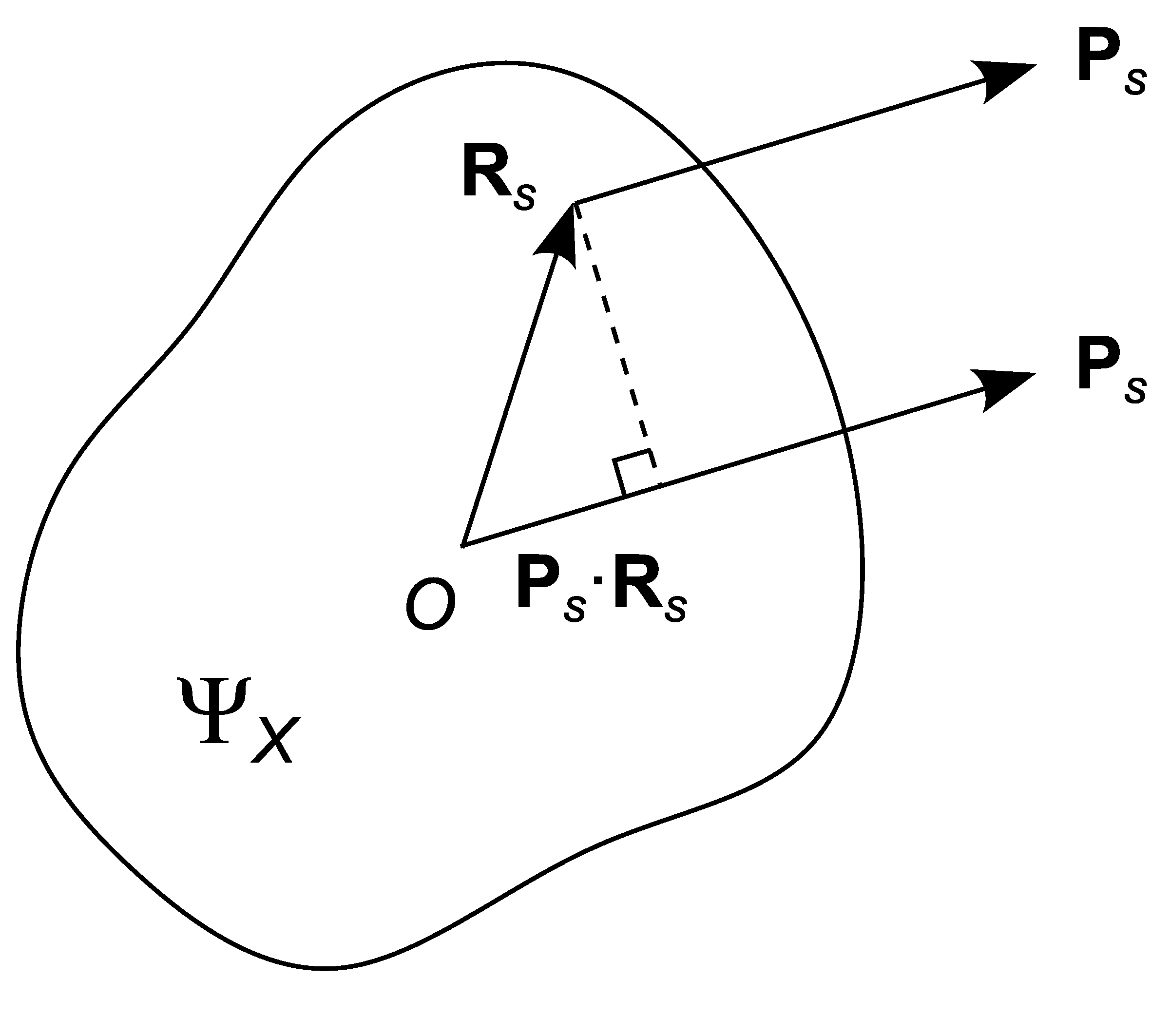}
      \caption{Schematic representation of the Fourier transform appeared in \cref{eq:another_nn}.
      This is an analogy with the atomic scattering factor, which appears in diffraction theory.\cite{cullity2001elements,waseda2011scattering}
      The plane waves having the amplitude $\Psi_X$ and the wave vector $\vb{P}_s$ are emitted from all positions of the target $s$th atom, and an interference occurs
      due to the superposition of these waves.
      $\vb{P}_s\cdot\vb{R}_s$ represents the phase difference between the waves emitted from the position $\vb{R}_s$ and the origin $O$.
      }\label{fig:squared_amplitude}
\end{figure}

The oscillation in the quantum momentum distribution for \ce{H2O(H)} can be understood qualitatively as follows. Substituting \cref{eq:tildepsix} into \cref{eq:nn=int_abspsi^2}, we can rewrite $n_s(\vb{P}_s)$ as
\begin{align}
    n_s(\vb{P}_s)=&
    \frac{1}{(2\uppi)^{3}}\int\dd\vb{R}_1\cdots\dd\vb{R}_{s-1}\dd\vb{R}_{s+1}\cdots\dd\vb{R}_N\notag\\
    &\times\abs{
    \int\Psi_X(\vb{R}_1,\dots,\vb{R}_N)
    \mathrm{e}^{-\mathrm{i}\vb{P}_s\cdot\vb{R}_s}\dd\vb{R}_s
    }^2.
    \label{eq:another_nn}
\end{align}
By analogy with the atomic scattering factor,\cite{cullity2001elements,waseda2011scattering} the Fourier transform in \cref{eq:another_nn} can be considered as a superposition of plane waves emitted from all positions $\vb{R}_s$ with amplitude $\Psi_X$ and wave vector $\vb{P}_s$, which is illustrated in \cref{fig:squared_amplitude}.
Then, the momentum distribution of the target atom corresponds to the intensity of the resultant wave averaged over all spatial configurations of the other atoms $(\vb{R}_1,\dots,\vb{R}_{s-1},\vb{R}_{s+1},\dots,\vb{R}_N)$.

The amplitude $\Psi_X$ of the plane wave, defined in the Cartesian coordinate system, is separated into translational and rotational-vibrational parts, as was the case for the generalized coordinate system:
\begin{align}
    \Psi_X(\vb{R}_1,\dots,\vb{R}_N)
    =\Psi_X^\mathrm{T}(\vb{R}_\mathrm{G})
    \Psi_X^\mathrm{RV}(\boldsymbol{\uprho}_1,\dots,\boldsymbol{\uprho}_N)
    \label{eq:PsiX=PsiXT*PsiXRV}
\end{align}
where $\Psi_X^\mathrm{T}$ is identical to $\Psi_u^\mathrm{T}$, and $\Psi_X^\mathrm{RV}$ is regarded as a function on the whole position space, satisfying
\begin{align}
    \Psi_X^\mathrm{RV}(\vb{R}_1,\dots,\vb{R}_N)=\Psi_X^\mathrm{RV}(\boldsymbol{\uprho}_1,\dots,\boldsymbol{\uprho}_N).
\end{align}
Since we assumed $n^\mathrm{T}=\delta(\vb{P}_\mathrm{T})$, $\Psi_X^\mathrm{T}$ can be expressed as the wavefunction of a free particle in its ground state:
\begin{align}
    \Psi_X^\mathrm{T}(\vb{R}_\mathrm{G})
    =\lim_{V_\mathrm{G}\to\infty}\frac{1}{\sqrt{V_\mathrm{G}}},
    \label{eq:PsiXT=const.}
\end{align}
where $V_\mathrm{G}$ is the volume of the space $\vb{R}_\mathrm{G}$.
Therefore, $\Psi_X$ is independent of $\vb{R}_\mathrm{G}$ and is determined only by the rotatinal-vibrational part $\Psi_X^\mathrm{RV}$.

In the vibrational ground state, $\Psi_X^\mathrm{RV}$ has an amplitude only near equilibrium atomic configurations $(\boldsymbol{\uprho}_1,\dots,\boldsymbol{\uprho}_N)\simeq(\boldsymbol{\uprho}_1^0,\dots,\boldsymbol{\uprho}_N^0)$ of the target molecule. 
If we treat the molecule as a perfect rigid rotor, only the orientational average ($\int\sin\theta\dd\Theta$) of $(\boldsymbol{\uprho}_1^0,\dots,\boldsymbol{\uprho}_{s-1}^0,\boldsymbol{\uprho}_{s+1}^0,\dots,\boldsymbol{\uprho}_N^0)$ needs to be considered in order to take into account the spatial average of $(\vb{R}_1,\dots,\vb{R}_{s-1},\vb{R}_{s+1},\dots,\vb{R}_N)$ in \cref{eq:another_nn}.
As a result of this rigid rotor model, the atomic momentum distribution can be expressed as
\begin{align}
    n_s(\vb{P}_s)
    \propto
    &\int\sin\theta\dd\Theta\left|
    \int_D\dd\vb{R}_s\,\mathrm{e}^{-\mathrm{i}\vb{P}_s\cdot\vb{R}_s}\right.\notag\\
    &\left.\vphantom{\int\dd\vb{R}_s\,\mathrm{e}^{-\mathrm{i}\vb{P}_s\cdot\vb{R}_s}}
    \times\Psi_X^\mathrm{RV}(\boldsymbol{\uprho}_1^0,\dots,\vb{R}_{s},\dots,\boldsymbol{\uprho}_N^0) 
    \right|^2.
    \label{eq:nn_rigidrotor}
\end{align}
Here the region $D$ indicates not only the initial configuration $\vb{R}_s=\boldsymbol{\uprho}_s^0$ but also the other configurations of $\vb{R}_s\neq\boldsymbol{\uprho}_s^0$ that coincide with the equilibrium structure. The presence of the $\vb{R}_s\neq\boldsymbol{\uprho}_s^0$ thus means that the target atom is spatially delocalized, resulting in interference of the plane waves emitted from them.

\begin{figure}[ht]
  \centering  \includegraphics{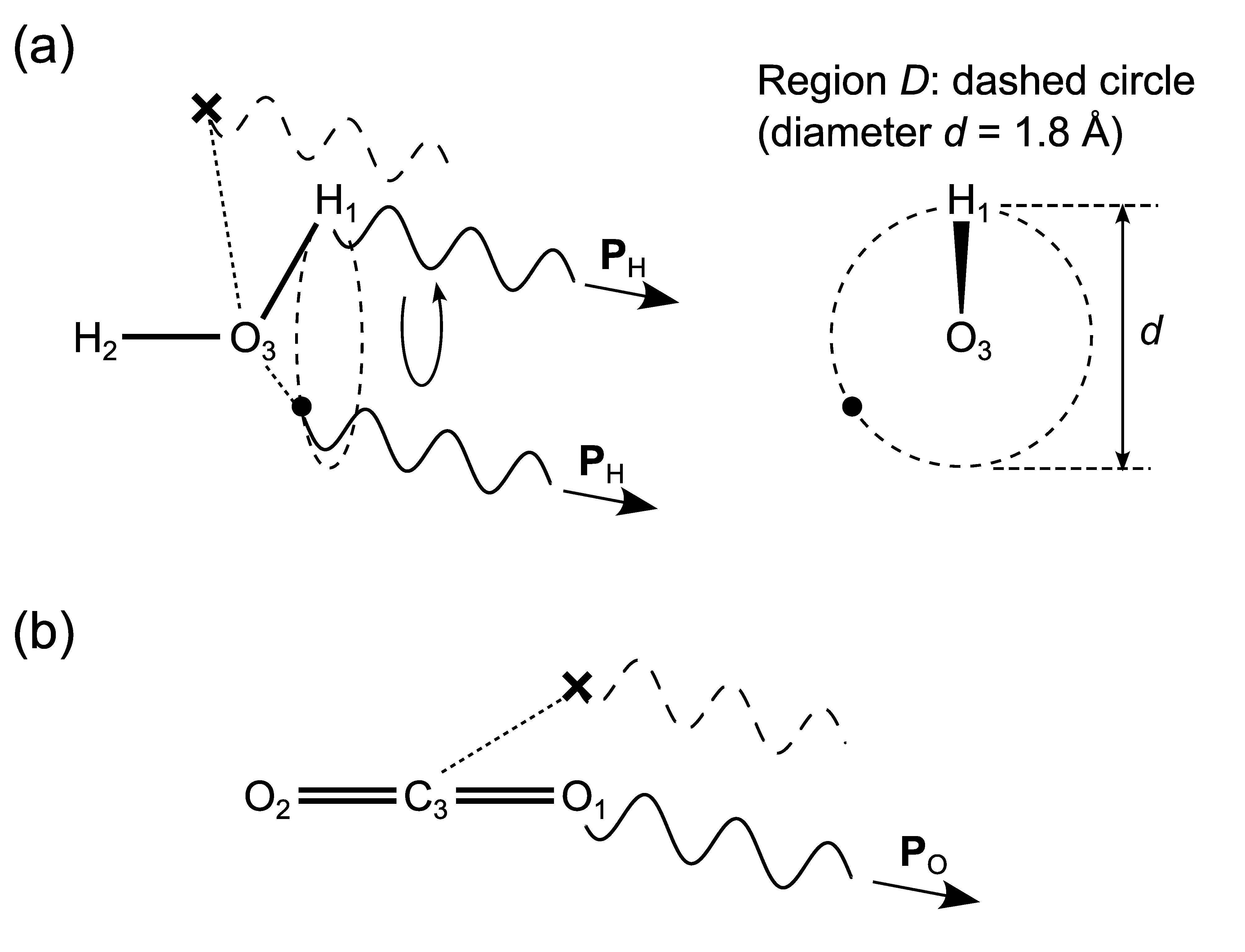}
  \caption{The plane waves with a specific $\vb{P}_s$ emitted from the target atom in the rigid rotor model: 
  (a) Plane waves can be emitted from any point on the dashed circle $D$ for \ce{H1} atom in \ce{H2O}, and hence they can interfere, while (b) only a single wave can be emitted from \ce{O1} atom in \ce{CO2}.} 
  \label{fig:np_model}
\end{figure}

\cref{fig:np_model} illustrates the interference effect on the $n_s(\vb{P}_s)$ in terms of the spatial delocalization of the target atom. 
The Fourier integral in \cref{eq:another_nn} is carried out with respect to the $\vb{R}_s$, while keeping the other atoms fixed at a configuration $(\vb{R}_1,\dots,\vb{R}_{s-1},\vb{R}_{s+1},\dots,\vb{R}_N)$.
For the $\vb{R}_s$ region where $\Psi_X(\vb{R}_1,\dots,\vb{R}_N)\simeq0$, the integrand is vanishingly small.
In the case of \ce{H2O}, thereby, \ce{H1} configuration which leads to significantly deformed structures from the equilibrium one [e.g., a configuration marked by cross in \cref{fig:np_model}(a)] have no contribution to the Fourier transform in \cref{eq:another_nn}.
However, the \ce{H1} configurations which are identical to those obtained by the rotation with respect to the \ce{H2-O3} axis can have some contributions to the integral.
The dashed circle drawn in \cref{fig:np_model}(a) indicates such possible configurations of \ce{H1}, which corresponds to the region $D$ specified in \cref{eq:nn_rigidrotor} for the rigid rotor model.
Quantum mechanically, this situation represents rotational delocalization of the target \ce{H1} atom with respect to the \ce{H2} and \ce{O3} atoms.
Therefore, the plane waves emitted from the different positions $\vb{R}_\mathrm{H}$ of the \ce{H1} interfere constructively or destructively each other depending on the momentum $\vb{P}_\mathrm{H}$ of the \ce{H1}, which results in the oscillation in the proton momentum distribution. 
This kind of interference can occur in \ce{H2O} because of the nonlinear molecular structure.
For the linear triatomic molecule, \ce{CO2}, such interference is impossible since the \ce{O1} atom is localized with respect to the \ce{O2} and \ce{C3} atoms, and hence no oscillation due to the quantum interference appears in the O atom's momentum distribution in \ce{CO2}. 

\begin{figure}[ht]
  \centering  \includegraphics{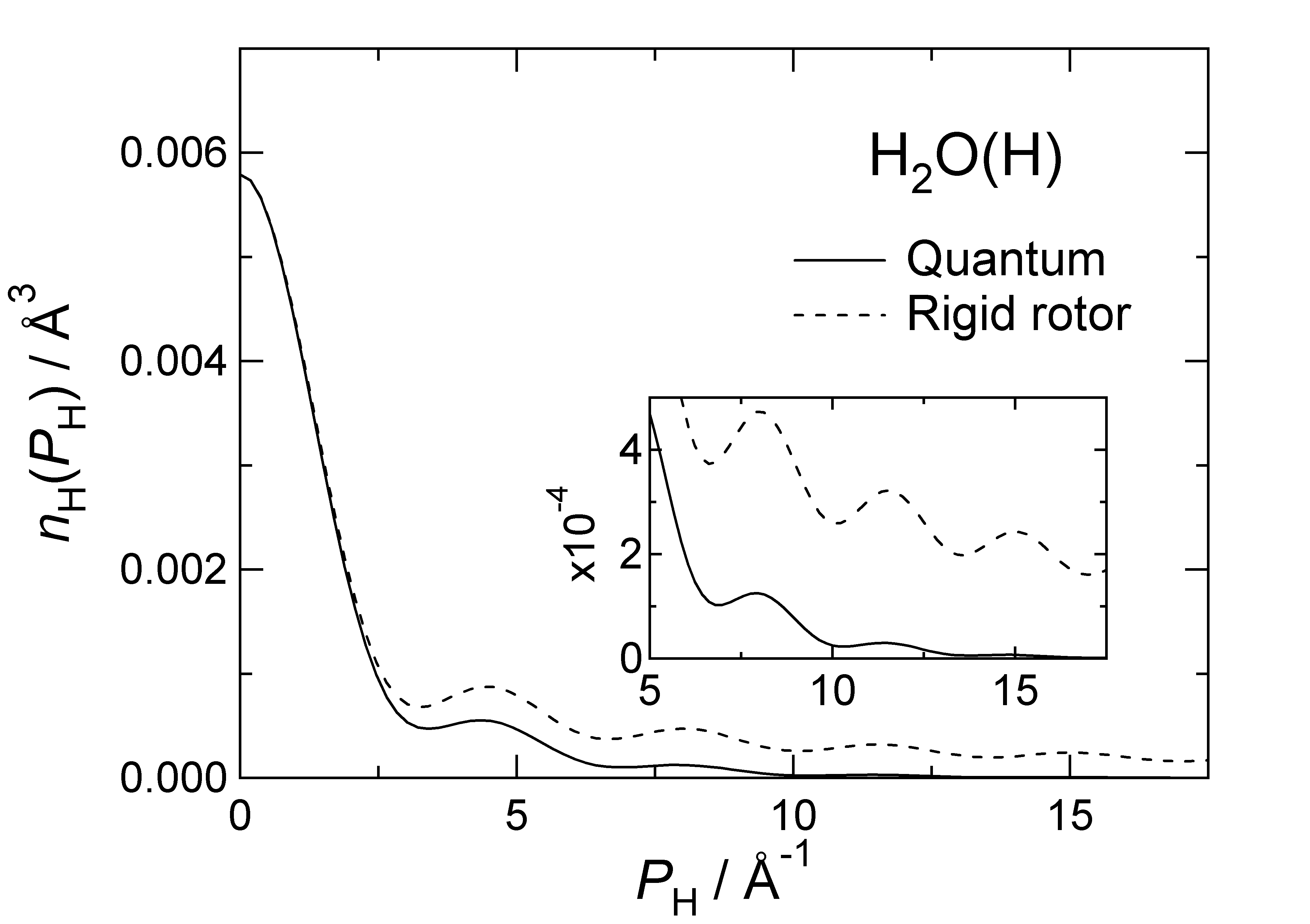}
  \caption{The proton momentum distributions $n_\mathrm{H}(P_\mathrm{H})$ of \ce{H2O} in the ground state. The solid and dashed lines represent the quantum and rigid rotor model calculations, respectively.}\label{fig:rigid_rotor_model_calculation}
\end{figure}

To demonstrate the validity of the above interpretation, we performed the calculation of \cref{eq:nn_rigidrotor} for \ce{H2O(H)} using the rigid rotor model.
The region of integration $D$ is set to be just a circumference with a diameter $d=\SI{1.8}{\angstrom}$ as shown in \cref{fig:np_model}, which is formed by the trajectory of the target \ce{H1} atom rotating with respect to the \ce{H2-O3} axis by keeping the equilibrium structure of \ce{H2O}. 
In addition, $\Psi_X^\mathrm{RV}$ is just a constant on the $D$ for the rotational ground state.
Using these simplifications, we can rewrite \cref{eq:nn_rigidrotor} as
\begin{align}
    n_\mathrm{H}(P_\mathrm{H})
    \propto
    \int_0^{\uppi}\sin\theta\dd\theta
    \abs{J_0\left(
    P_\mathrm{H}d\sin\theta/2
    \right)}^2,
    \label{eq:nH_rigid_rotor}
\end{align}
where $J_0$ is the zeroth-order Bessel function of the first kind.
\cref{fig:rigid_rotor_model_calculation} compares the rigid rotor model calculation of \cref{eq:nH_rigid_rotor} with the quantum calculation obtained from \cref{eq:nP_cylindrical}.
It is evident from the figure that the period of the oscillation is well reproduced by the restricted rigid rotor model though the intensity is overestimated.
Therefore, it is concluded that the essential feature of the oscillation is originated from the quantum interference due to the rotational delocalization of the target atom. 

\begin{figure}[ht]
  \includegraphics{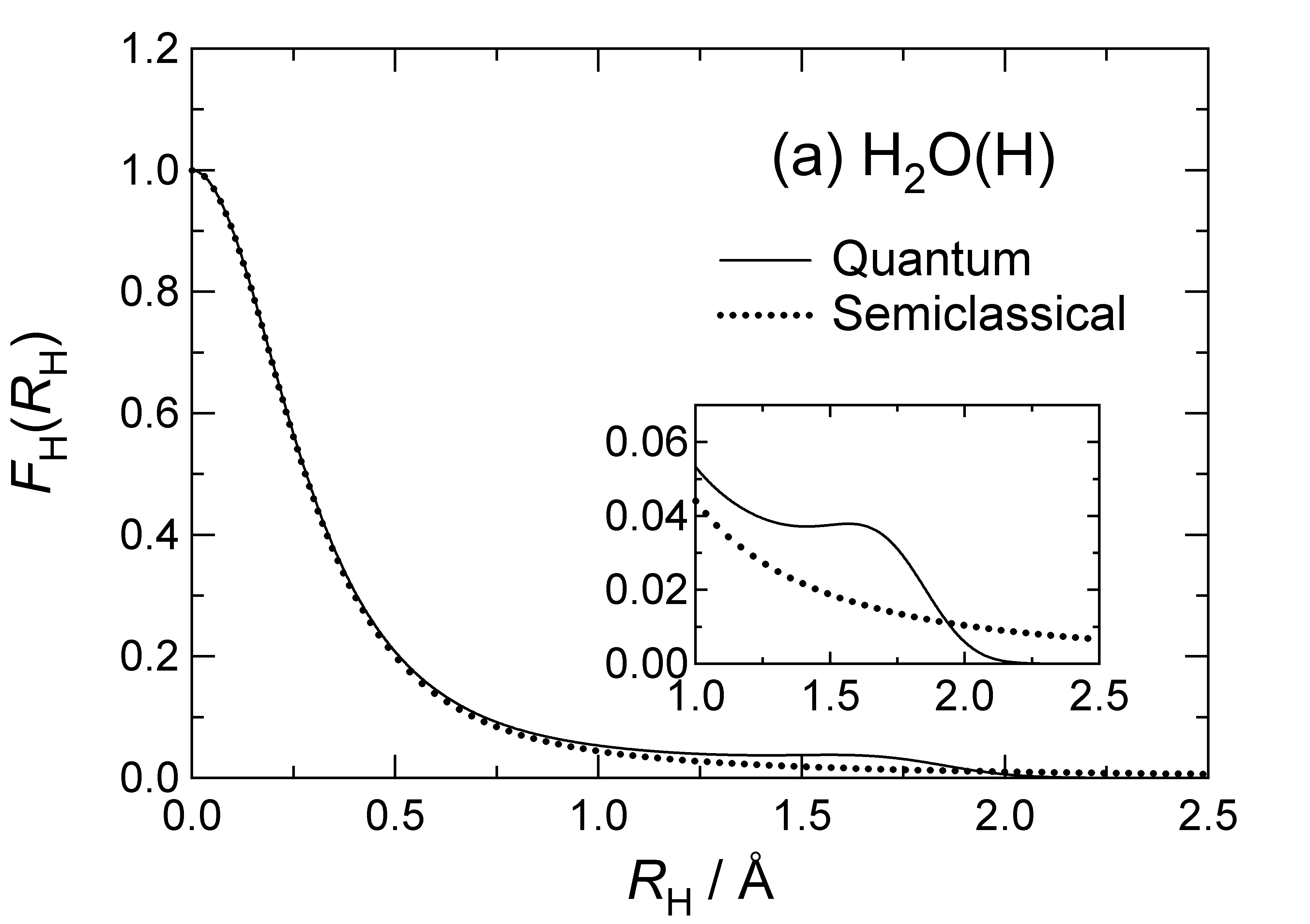}
  \includegraphics{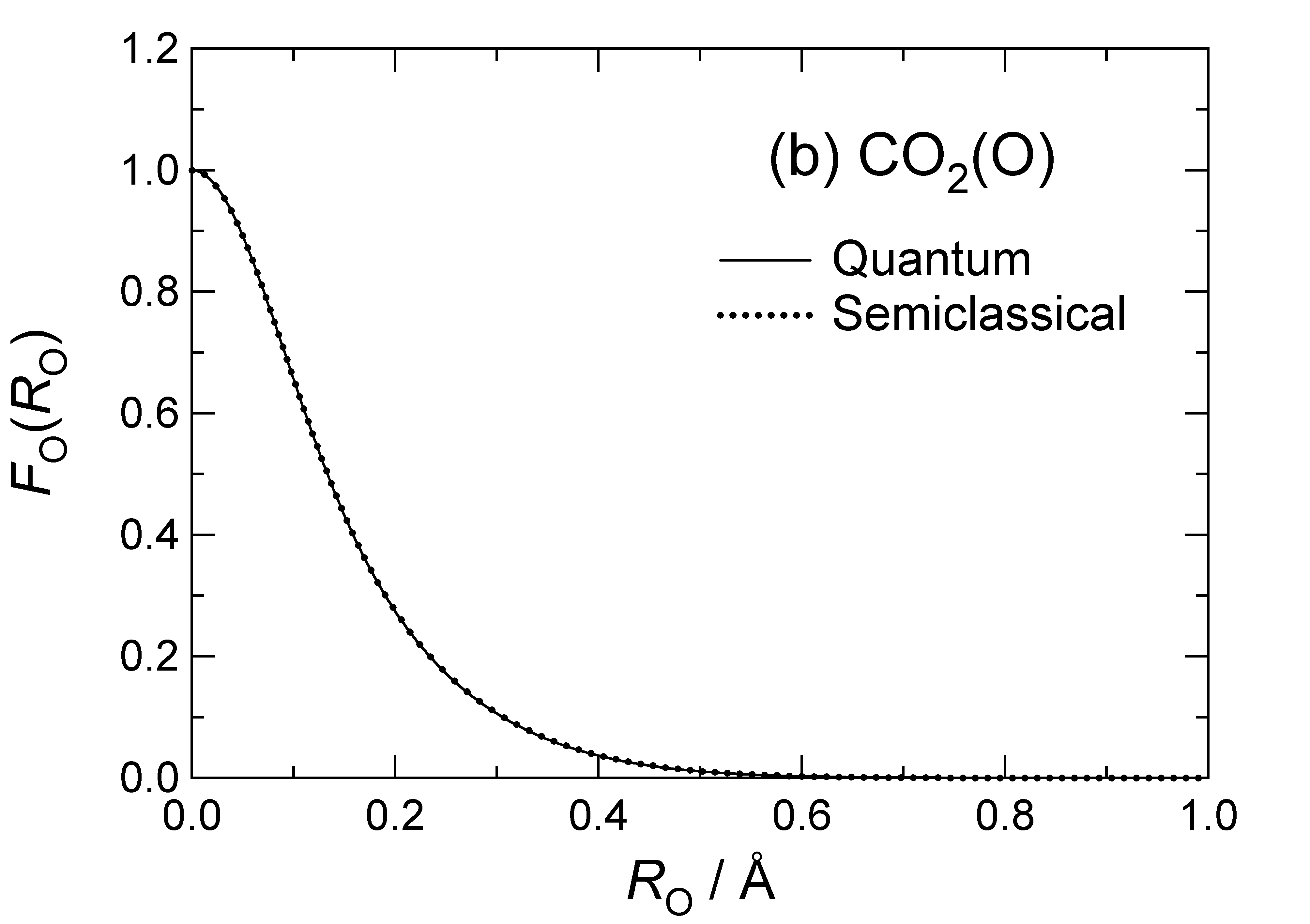}
  \caption{The inverse Fourier transform of the atomic momentum distributions $F_s(R_s)$ in the ground state: 
  (a) H atom in \ce{H2O} and (b) O atom in \ce{CO2}. 
  The solid and dotted lines represent the quantum and semiclassical $(T=\SI{0}{K})$ calculations, respectively.}\label{fig:Fs}
\end{figure}

An alternative way of interpreting the oscillation in the quantum momentum distribution for \ce{H2O}(H) can be provided by the autocorrelation $F_s$ of the position-space wavefunction,
\begin{align}
    F_s(\vb{R}_s)
    =&\int\dd\vb{R}'_1\cdots\dd\vb{R}'_N\,\Psi_X^*(\vb{R}'_1,\dots,\vb{R}'_s,\dots,\vb{R}'_N)\notag\\
    &\times
    \Psi_X(\vb{R}'_1,\dots,\vb{R}'_s+\vb{R}_s,\dots,\vb{R}'_N),
    \label{eq:Fn_autocorrelation}
\end{align}
which can be obtained by the inverse Fourier transform $F_s$ of the atomic momentum distribution, 
\begin{align}
    F_s(\vb{R}_s)
    =\int n_s(\vb{P}_s)
    \mathrm{e}^{\mathrm{i}\vb{P}_s\cdot\vb{R}_s}\dd\vb{P}_s.
    \label{eq:Fn_definition}
\end{align}
Since we consider here only the rotational ground state, $n_s$ is isotropic. Using the expansion of a plane wave in spherical harmonics, $F_s$ is expressed as 
\begin{align}
    F_s(R_s)=4\uppi\int j_0(P_sR_s)n_s(P_s)P_s^2\dd P_s
    \label{eq:isotropic_Fn},
\end{align}
where $j_0$ is the zeroth-order spherical Bessel function.
\cref{fig:Fs} shows the inverse Fourier transform of the momentum distributions of \ce{H2O}(H) and \ce{CO2}(O) calculated by \cref{eq:isotropic_Fn}.
For comparison, also shown are the results calculated by using semiclassical $n_s(P_s)$ for \cref{eq:isotropic_Fn}.
In the case of \ce{H2O}(H), in the small $R_\mathrm{H}$ region, the quantum and semiclassical results are almost identical to each other.
However, as $R_\mathrm{H}$ increases, there is a relevant deviation of the quantum result from the semiclassical one, as can be clearly seen in the insert of \cref{fig:Fs}(a). 
In particular, the result of the quantum calculation exhibits a sharp decrease to zero at around \SI{1.8}{\angstrom}.
In the case of \ce{CO2}(O), there is no such a sharp decay; there is no appreciable difference in $F_\mathrm{O}(R_\mathrm{O})$ between the quantum and semiclassical results. 

As mentioned in \cref{Computational Details} in detail, the only difference between the quantum and semiclassical calculations is the treatment of rotational motion. In the small $R_s$ region, the autocorrelation is dominated by the overlap of the vibrational wavefunctions between the original and displaced positions. In this region, the semiclassical results are in good agreement with the quantum ones, as the semiclassical calculation can take into account the quantum-mechanical vibrational autocorrelation. Therefore, the difference in the $F_\mathrm{H}$ between the quantum and semiclassical results of \ce{H2O}(H) demonstrates clearly the quantum effect of autocorrelation due to rotational motion, i.e., quantum rotational autocorrelation.

The sharp drop in the $F_\mathrm{H}$ of \ce{H2O}(H), which does not appear in the semiclassical result, can be interpreted by the behavior of the autocorrelation in the rigid rotor model. 
To derive this model, we eliminate the effect of the translational motion in \cref{eq:Fn_autocorrelation} by factorizing the autocorrelation function into the translational and rotational-vibrational parts.\cite{colognesi2001deep}
Substituting \cref{eq:PsiX=PsiXT*PsiXRV} into  
\cref{eq:Fn_autocorrelation} (or \cref{eq:nn=nT*nRV} into \cref{eq:Fn_definition}), we obtain
\begin{align}
    F_s(\vb{R}_s)
    =F^\mathrm{T}\left(\frac{M_s}{M_\mathrm{T}}\vb{R}_s\right)
    F_s^\mathrm{RV}(\vb{R}_s),
    \label{eq:Fn=FT*FRV}
\end{align}
where
\begin{align}
    &F^\mathrm{T}(\vb{R}_\mathrm{G})
    =\int n^\mathrm{T}(\vb{P}_\mathrm{T})
    \mathrm{e}^{\mathrm{i}\vb{P}_\mathrm{T}\cdot\vb{R}_\mathrm{G}}\dd\vb{P}_\mathrm{T}\notag\\
    &\hphantom{F^\mathrm{T}(\vb{R}_\mathrm{G})}
    =\int\Psi_X^{\mathrm{T}*}(\vb{R}'_\mathrm{G})
    \Psi_X^{\mathrm{T}}(\vb{R}'_\mathrm{G}+\vb{R}_\mathrm{G})\dd\vb{R}'_\mathrm{G},
    \label{eq:FT}\\
    &F_s^\mathrm{RV}(\boldsymbol{\uprho}_s)
    =\int n^\mathrm{RV}_s(\boldsymbol{\uppi}_s)
    \mathrm{e}^{\mathrm{i}\boldsymbol{\uppi}_s\cdot\boldsymbol{\uprho}_s}\dd\boldsymbol{\uppi}_s\notag\\
    &\hphantom{F_s^\mathrm{RV}(\boldsymbol{\uprho}_s)}
    =\int_V\dd\boldsymbol{\uprho}'_1\cdots\dd\boldsymbol{\uprho}'_N
    \,\Psi_X^{\mathrm{RV}*}(\boldsymbol{\uprho}'_1,\dots,\boldsymbol{\uprho}'_s,\dots,\boldsymbol{\uprho}'_N)\notag\\
    &\hphantom{F_s^\mathrm{RV}(\boldsymbol{\uprho}_s)=}
    \times\Psi_X^{\mathrm{RV}}(\boldsymbol{\uprho}'_1,\dots,\boldsymbol{\uprho}'_s+\boldsymbol{\uprho}_s,\dots,\boldsymbol{\uprho}'_N)
    \label{eq:FnRV}
\end{align}
with the position space $V$ satisfying $\sum_iM_i\boldsymbol{\uprho}_i=0$.
From \cref{eq:FT}, 
$n^\mathrm{T}=\delta(\vb{P}_\mathrm{T})$ yields $F^\mathrm{T}=1$, which is consistent with \cref{eq:PsiXT=const.}.
Thus, \cref{eq:Fn=FT*FRV} becomes
\begin{align}
    F_s(\vb{R}_s)=F_s^\mathrm{RV}(\vb{R}_s).
\end{align}
In the same way as for the atomic momentum distribution, the rigid rotor model gives the autocorrelation function as
\begin{align}
    F_s(\vb{R}_s)
     \propto&\int\sin\theta\dd\Theta\,\Psi_X^{\mathrm{RV}*}(\boldsymbol{\uprho}_1^0,\dots,\boldsymbol{\uprho}^0_s,\dots,\boldsymbol{\uprho}^0_N)\notag\\
    &\times\Psi_X^{\mathrm{RV}}(\boldsymbol{\uprho}^0_1,\dots,\boldsymbol{\uprho}^0_s+\vb{R}_s,\dots,\boldsymbol{\uprho}^0_N).
\end{align}

\begin{figure}[ht]
  \centering
  \includegraphics{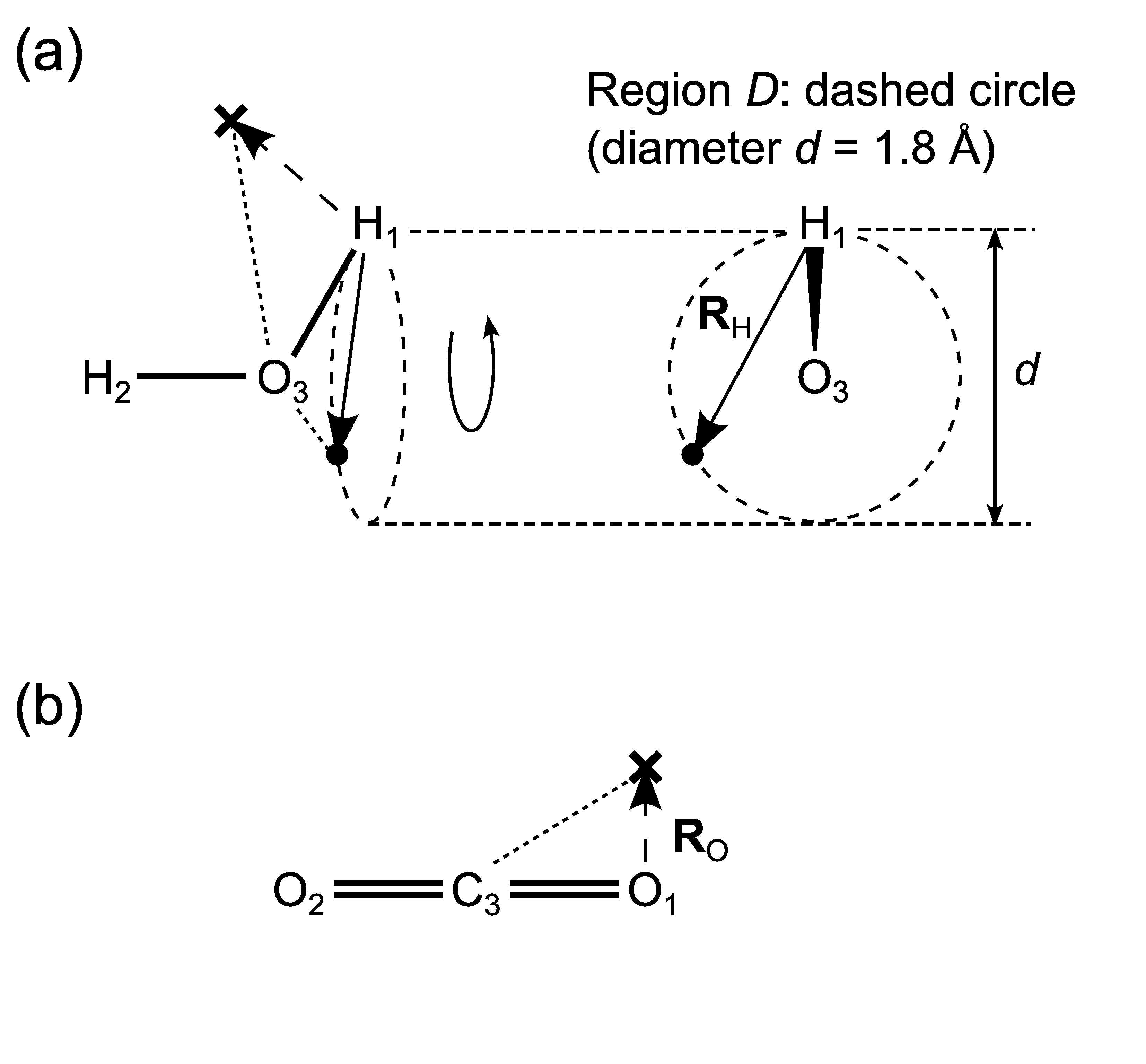}
  \caption{Illustrative examples of possible displacements of the target atom within the rigid rotor model. (a) the position autocorrelation of \ce{H2O(H)} remains when the \ce{H1} atom rotates with respect the \ce{H2-O3} axis, while (b) that of \ce{CO2(O)} vanishes for any displacement of the \ce{O1} atom.}\label{fig:Fs_model}
\end{figure}

In general, as $R_s$ increases, $(\boldsymbol{\uprho}^0_1,\dots,\boldsymbol{\uprho}^0_s+\vb{R}_s,\dots,\boldsymbol{\uprho}^0_N)$ deviates from the initial equilibrium configuration of the atoms, and hence the autocorrelation value decreases monotonically due to the decrease of the amplitude of the vibrational wavefunctions around $\boldsymbol{\uprho}^0_s$.
However, when a displaced atomic configuration $\boldsymbol{\uprho}^0_s+\vb{R}_s$ accidentally coincides with the equilibrium molecular structure, the autocorrelation will retain a certain value. 
For \ce{H2O}, this situation is illustrated in \cref{fig:Fs_model}(a).
Quantum-mechanically, all the configurations in which the \ce{H1} atom is on the dashed circle in \cref{fig:Fs_model}(a) contribute to the autocorrelation coherently, since the \ce{H1} atom is delocalized over the circle.
On the other hand, there is no such rotational autocorrelation in the semiclassical calculation because the atomic momentum distribution within the semiclassical framework is obtained just by an orientational average (incoherent sum) of the atomic momentum distribution for one equilibrium configuration.\cite{colognesi2001deep,ivanov1967interaction}
Therefore, it is concluded that the difference between the semiclassical and quantum results of $F_\mathrm{H}$ for \ce{H2O}(H) is attributed to this rotational autocorrelation. 
It is noted that the upper limit of  $R_\mathrm{H}$ in the rotational autocorrelation is determined by the equilibrium molecular geometry of \ce{H2O}. 
It is $\SI{1.8}{\angstrom}$, the diameter $d$ on which the \ce{H1} atom is delocalized, corresponding to the position where the autocorrelation decays to zero in \cref{fig:Fs}(a). 
In the case of \ce{CO2}, on the other hand, no rotational autocorrelation occurs because the displacement of the target \ce{O1} atom always results in a different molecular structure from the stable linear structure, as shown in \cref{fig:Fs_model}(b).

The above discussion about the quantum delocalization effect on the atomic momentum distributions can be extended to any molecule, not just to ground-state triatomic molecules.
Another example is the oscillatory structure in the atomic momentum distribution of a diatomic molecule in the vibrational ground state. Since the oscillation is attributed to the spherical delocalization of the target atom around the other atom, diatomic molecules inherently exhibit the interference effect on the atomic momentum distribution.
As mentioned in the \cref{Introduction}, it was pointed out that the wavelength (or period) of the oscillation depends on the internuclear distance.\cite{colognesi1999deep} This point can be revisited in terms of delocalization effect of the target atom with respect to the other atom; the internuclear distance determines the size of the sphere from which the plane waves are emitted.

\section{Conclusion}
\label{Conclustion}
This study has reported the quantum theory for the atomic momentum distributions of polyatomic molecules, which is based on the momentum-space molecular wavefunctions in rotational-vibrational eigenstates. 
The rigid-rotor-harmonic-oscillator approximation has been developed to significantly reduce the computational cost, which enables one to extend the quantum-mechanical calculation to any kinds of isolated molecules in practice.

The proton momentum distribution in the ground-state \ce{H2O} was found to clearly show an oscillatory structure, or interference fringes.
This oscillation is originated purely from the quantum nature and hence is not appeared on the momentum distribution when the molecular rotation is treated as classical. 
On the other hand, no such oscillation occurs on the oxygen momentum distribution in the ground-state \ce{CO2} molecule even in the quantum calculation.

The rigid rotor model shows that whether or not the interference fringes appear in the atomic momentum distribution in a polyatomic molecule is determined by whether the target atom is delocalized or localized with respect to the other atoms.
For the vibrational ground state, where the rigid rotor model is applicable, this finding leads to the following general rule that predicts necessarily the presence of interference fringes without rigorous calculations.
Firstly, the positions of all atoms are fixed in an equilibrium configuration (initial configuration) of the molecule.
Next, the target atom is displaced with keeping the other atoms in the same positions as the initial ones, to find one or more equilibrium configurations different from the initial configuration.
Then, if the different equilibrium configurations are obtained, the interference structure may appear, otherwise no interference fringes are expected in the atomic momentum distribution.

Although a profound theoretical understanding for the delocalization effect on atomic momentum distributions has been gained for isolated molecules, experimental investigation has not been attempted so far. 
This is partially because the difficulty in a measurement on a single quantum state of dilute gas-phase molecules by using DINS, which has mostly been applied to condensed matter systems due to the limitation in sensitivity.
However, the recent development of electron-atom Compton scattering, or Atomic Momentum Spectroscopy,\cite{tachibana2022direct,vos2001observing,cooper2007quasielastic,vos2013elastic,yamazaki2017development,onitsuka_asymptotic_2022} would make it possible to measure the atomic momentum distribution in gas-phase molecules in their single quantum state.
A combination of a recently developed highly-sensitive apparatus\cite{yamazaki2017development} with adiabatic cooling would offer a promising means of studying the quantum effects on the atomic momentum distribution in a single molecule, and the observation of the oscillatory structure in a rovibrational state would become feasible in the near future.

\appendix*
\section{Simplification of the Jacobian}
\label{appendix}
\paragraph{Nonlinear molecules}
From \cref{eq:rn,eq:rhon=R(rho'n0+drho_n)}, the derivatives of the mass-weighted coordinates $(q_a)$ with respect to the generalized coordinates $(u_a)$ become
\begin{widetext}
\begin{align}
    \dd\left(M_i^{1/2}\vb{R}_i\right)
    &=M_i^{1/2}\left(
    \dd\vb{R}_\mathrm{G}
    +\dd R\vb{r}_i
    +R\dd\Delta\vb{r}_i
    \right)
    =M_i^{1/2}R\left(
    R^{-1}\dd\vb{R}_\mathrm{G}
    +R^{-1}\dd R\vb{r}_i
    +\dd\Delta\vb{r}_i
    \right)\notag\\
    &=R\left[
    M_i^{1/2}R^{-1}\dd\vb{R}_\mathrm{G}
    +\left(\vu{e}_\phi\dd\phi
      +\vu{e}_\theta\dd\theta
      +\vu{e}_\chi\dd\chi\right)\times M_i^{1/2}\vb{r}_i
    +\sum_{k=1}^{3N-6}\vb{l}_{i,k}\dd Q_k
    \right]
\end{align}
with $M_i^{1/2}\vb{R}_i=(q_{3i-2},q_{3i-1},q_{3i})$, and hence we can write the Jacobian matrix $(\partial q_a/\partial u_b
)$ as
\begin{align}
    \left(\pdv{q_a}{u_b}\right)
    =R_{3N}    
    \begin{pmatrix}
    M_1^{1/2}R^{-1}
    &\vu{e}_\phi\times M_1^{1/2}\vb{r}_1
    &\vu{e}_\theta\times M_1^{1/2}\vb{r}_1
    &\vu{e}_\chi\times M_1^{1/2}\vb{r}_1
    &\vb{l}_{1,1}&\vb{l}_{1,2}&\cdots&\vb{l}_{1,3N-6}\\
    M_2^{1/2}R^{-1}
    &\vu{e}_\phi\times M_2^{1/2}\vb{r}_2
    &\vu{e}_\theta\times M_2^{1/2}\vb{r}_2
    &\vu{e}_\chi\times M_2^{1/2}\vb{r}_2
    &\vb{l}_{2,1}&\vb{l}_{2,2}&\cdots&\vb{l}_{2,3N-6}\\
    \vdots&\vdots&\vdots&\vdots&\vdots&\vdots&\ddots&\vdots\\
    M_N^{1/2}R^{-1}
    &\vu{e}_\phi\times M_N^{1/2}\vb{r}_N
    &\vu{e}_\theta\times M_N^{1/2}\vb{r}_N
    &\vu{e}_\chi\times M_N^{1/2}\vb{r}_N
    &\vb{l}_{N,1}&\vb{l}_{N,2}&\cdots&\vb{l}_{N,3N-6}
\end{pmatrix},
\end{align}
\end{widetext}
where all bold symbols represent column vectors, and
the directions of the infinitesimal rotation vector $\vu{e}_\phi$, $\vu{e}_\theta$, and $\vu{e}_\chi$ are given by
\begin{subequations}
\begin{align}
    \vu{e}_\phi&=
    \begin{pmatrix}
        -\sin\theta\cos\chi&
        \sin\theta\sin\chi&
        \cos\theta
    \end{pmatrix}
    ^\top,\\
    \vu{e}_\theta&=
    \begin{pmatrix}
        \sin\chi&
        \cos\chi&
        0
    \end{pmatrix}
    ^\top,\\
    \vu{e}_\chi&=
    \begin{pmatrix}
        0&0&1
    \end{pmatrix}
    ^\top,
\end{align}    
\end{subequations}
and $R_{3N}=\bigoplus_{i=1}^NR$ is the direct sum of $N$ rotation matrices in three dimensions $R$.
Using \cref{eq:condition_of_l} and the translational Eckart conditions
\begin{align}
    \sum_iM_i\vb{r}_i=0,
\end{align}
we can simplify the determinant $g_u$ of \cref{eq:g_uij} to
\begin{align}
    g_u
    &=\abs{\left(\pdv{q_c}{u_a}\right)^\top
    \left(\pdv{q_c}{u_b}\right)}
    =
    \begin{vmatrix}
    M_\mathrm{T}E_3&&\\
    &A&B^\top\\
    &B&E_{3N-6}
    \end{vmatrix}\notag\\
    &=M_\mathrm{T}^3\det(A-B^\top B),
    \label{eq:g_simplification}
\end{align}
where $E_n$ is the identity matrix of size $n$, and the $3\times3$ matrix $A$ and the $(3N-6)\times3$ matrix $B$ are written as
\begin{align}
    &A=T^\top(I_{\alpha\beta})T
    \label{eq:matrixA},\\
    &B^\top=T^\top\sum_k
    \begin{pmatrix}
        \boldsymbol{\upzeta}_{k1}&
        \boldsymbol{\upzeta}_{k2}&
        \cdots&
        \boldsymbol{\upzeta}_{k,3N-6}
    \end{pmatrix}Q_k
    \label{eq:matrixB}
\end{align}
with the $3\times3$ matrix
$T=
\begin{pmatrix}
    \vu{e}_\phi&\vu{e}_\theta&\vu{e}_\chi
\end{pmatrix}$
and
$\boldsymbol{\upzeta}_{km}=
\begin{pmatrix}
    \zeta_{km}^x&\zeta_{km}^y&\zeta_{km}^z
\end{pmatrix}^\top$.
From \cref{eq:matrixA,eq:matrixB}, we obtain
\begin{align}
    A-B^\top B
    =
    T^\top(I'_{\alpha\beta})T
    \label{eq:A-B@B}
\end{align}
Substituting \cref{eq:A-B@B} into \cref{eq:g_simplification}, $g_u$ is rewritten as
\begin{align}
    g_u=M^3_\mathrm{T}\det(I'_{\alpha\beta})\sin^2\theta,
\end{align}
where we use $\det(T)=\det(T^\top)=-\sin\theta$.
Calculating \cref{eq:g_Xij} from $(q_a)=(M_1^{1/2}\vb{R}_1,\dots,M_N^{1/2}\vb{R}_N)$, the determinant $g_X$ is
\begin{align}
    g_X=\prod_{i}M_i^3.
\end{align}
Therefore, $\abs{|J|}=\sqrt{g_{u}/g_{X}}$ can be expressed as
\begin{align}
    \abs{|J|}&=
    \left[\frac{M_\mathrm{T}^3}{\prod_iM_i^3}\det(I'_{\alpha\beta})\right]^{1/2}\sin\theta
    \equiv G^{1/2}\sin\theta.
\end{align}
\paragraph{Linear molecules}
As mentioned above, for linear molecules, the Euler angle $\chi$ in the rotation matrix is set as $\chi=0$. Thus, the derivatives of $(q_a)$ are
\begin{widetext}
\begin{align}
    \dd\left(M_i^{1/2}\vb{R}_i\right)
    =R\left[
    M_i^{1/2}R^{-1}\dd\vb{R}_\mathrm{G}
    +\left(\vu{e}_\phi\dd\phi
      +\vu{e}_\theta\dd\theta
      \right)\times M_i^{1/2}\vb{r}_i
    +\sum_{k=1}^{3N-5}\vb{l}_{i,k}\dd Q_k
    \right],
\end{align}
which yield
\begin{align}
    \left(\pdv{q_a}{u_b}\right)
    =R_{3N}    
    \begin{pmatrix}
    M_1^{1/2}R^{-1}
    &\vu{e}_\phi\times M_1^{1/2}\vb{r}_1
    &\vu{e}_\theta\times M_1^{1/2}\vb{r}_1
    &\vb{l}_{1,1}&\vb{l}_{1,2}&\cdots&\vb{l}_{1,3N-5}\\
    M_2^{1/2}R^{-1}
    &\vu{e}_\phi\times M_2^{1/2}\vb{r}_2
    &\vu{e}_\theta\times M_2^{1/2}\vb{r}_2
    &\vb{l}_{2,1}&\vb{l}_{2,2}&\cdots&\vb{l}_{2,3N-5}\\
    \vdots&\vdots&\vdots&\vdots&\vdots&\ddots&\vdots\\
    M_N^{1/2}R^{-1}
    &\vu{e}_\phi\times M_N^{1/2}\vb{r}_N
    &\vu{e}_\theta\times M_N^{1/2}\vb{r}_N
    &\vb{l}_{N,1}&\vb{l}_{N,2}&\cdots&\vb{l}_{N,3N-5}
    \end{pmatrix}.
\end{align}    
\end{widetext}
In this case, $\vu{e}_\phi$ and  $\vu{e}_\theta$ are written as
\begin{subequations}
\begin{align}
    \vu{e}_\phi&=
    \begin{pmatrix}
        -\sin\theta&
        0&
        \cos\theta
    \end{pmatrix}
    ^\top,\\
    \vu{e}_\theta&=
    \begin{pmatrix}
        0&
        1&
        0
    \end{pmatrix}
    ^\top.
\end{align}
\end{subequations}
In the same way as for nonlinear molecules, we obtain
\begin{align}
    g_u
    &=\abs{\left(\pdv{q_c}{u_a}\right)^\top
    \left(\pdv{q_c}{u_b}\right)}
    =
    \begin{vmatrix}
    M_\mathrm{T}E_3&&\\
    &A&B^\top\\
    &B&E_{3N-5}
    \end{vmatrix}
    \notag\\
    &=M_\mathrm{T}^3\det(A-B^\top B)
    \label{eq:g_simplification(linear)}
\end{align}
where the $2\times2$ matrix $A$ and the $(3N-5)\times2$ matrix $B$ is given by
\begin{align}
    &A=T^\top(I_{\alpha\beta})T,
    \label{eq:matrixA(linear)}\\
    &B^\top=T^\top\sum_k
    \begin{pmatrix}
        \boldsymbol{\upzeta}_{k1}&
        \boldsymbol{\upzeta}_{k2}&
        \cdots&
        \boldsymbol{\upzeta}_{k,3N-5}
    \end{pmatrix}Q_k
    \label{eq:matrixB(linear)}
\end{align}
with the $3\times2$ matrix
$T=
\begin{pmatrix}
    \vu{e}_\phi&\vu{e}_\theta
\end{pmatrix}$.
From \cref{eq:matrixA(linear),eq:matrixB(linear)}, we have
\begin{align}
    A-B^\top B
    &=T^\top(I_{\alpha\beta}')T
    =
    \begin{pmatrix}
        I'\sin^2\theta&\\
        &I'
    \end{pmatrix},
    \label{eq:A-B@B(linear)}
\end{align}
where we use \cref{eq:inertia_tensor&momentum}.
Substituting \cref{eq:A-B@B(linear)} into \cref{eq:g_simplification(linear)}, we can rewrite $g_u$ as
\begin{align}
    g_u=M^3_\mathrm{T}I^{\prime2}\sin^2\theta.
\end{align}
Therefore, $\abs{|J|}$ for linear molecules can be expressed as
\begin{align}
    \abs{|J|}
    &=\left[\frac{M_\mathrm{T}^3}{\prod_i M_i^3}I'^2\right]^{1/2}\sin\theta
    \equiv G^{1/2}\sin\theta.
\end{align}

%
%

%

\begin{acknowledgments}
This work was partially supported by JSPS KAKENHI Grant Number JP20H02688. It was also supported in part by the Inamori Research Grants, the Research Foundation for Opto-Science and Technology, the MATSUO FOUNDATION, and the Institute for Quantum Chemical Exploration. This work was also partially supported by JST FOREST Program (Grant Number JPMJFR201X, Japan). 
\end{acknowledgments}

\bibliography{main}

\end{document}